\documentclass[12pt,preprint]{aastex}

\begin{document}

\title{MV Lyrae in Low, Intermediate and High States\footnotemark[1]}
\footnotetext[1]
{Based on observations made with the NASA/ESA Hubble Space Telescope, obtained at the
Space Telescope Science Institute, which is operated by the Association of Universities 
for Research in Astronomy, Inc. under NASA contract NAS 5-26555 and with the Apache Point 
Observatory 3.5 m telescope, which is operated by the Astrophysical Research Corporation.}


\author{Albert P. Linnell$^1$, Paula Szkody$^2$, Boris G\"{a}nsicke$^3$, Knox S. Long$^4$,
Edward M.Sion$^5$, D.W.Hoard$^6$, and Ivan Hubeny$^7$}
\affil{$^1$Department of Astronomy, University of Washington, Box 351580, Seattle,
WA 98195-1580\\
$^2$Department of Astronomy, University of Washington, Box 351580, Seattle, 
WA 98195-1580\\
$^3$Department of Physics, University of Warwick, CV4 7AL,
Coventry, UK\\
$^4$Space Telescope Science Institute, 3700 San Martin Drive, Baltimore, MD 21218\\
$^5$Department of Astronomy and Astrophysics, Villanova University,
Villanova, PA 19085\\
$^6$Spitzer Science Center, California Institute of Technology, Mail Code 220-6,
1200 E. California Blvd., Pasadena, CA 91125\\
$^7$NOAO,950 N. Cherry, Tucson,AZ 85726 and Steward Observatory and Department of Astronomy,
University of Arizona, Tucson, AZ 85721\\}
\email{$^1$linnell@astro.washington.edu\\
$^2$szkody@astro.washington.edu\\
$^3$boris.gaensicke@warwick.ac.uk\\
$^4$long@stsci.edu\\
$^5$edward.sion@villanova.edu\\
$^6$hoard@ipac.caltech.edu\\
$^7$hubeny@noao.edu or hubeny@as.arizona.edu\\}

\begin{abstract}

Archival {\it IUE} spectra of the VY Scl system MV Lyr, taken during an intermediate state,
can be best fit by an isothermal accretion disk extending half way to the tidal
cutoff radius. 
In contrast, a recent HST spectrum,
while MV Lyr was in a high state, can be best fit 
with a standard $T(R)$ profile for an accretion disk extending from an inner truncation
radius to an intermediate radius with an isothermal accretion disk beyond.
These fits use component star parameters determined from a study of MV Lyr in a low state. 
Model systems containing accretion disks with standard $T(R)$ profiles 
have continua that 
are too blue.
The observed high state absorption line spectrum exhibits excitation higher than
provided by the $T(R)$ profile, indicating likely line formation in a high temperature
region extending vertically above the accretion disk.
The
absorption lines show a blue shift and line broadening corresponding to formation 
in a low velocity
wind apparently coextensive with the high temperature region. Lines of N V, Si IV, C IV, and He II 
are anomalously
strong relative to our synthetic spectra, indicating possible composition effects,
but unmodeled excitation effects could also produce the anomalies.

An analysis of a low state of MV Lyrae, considered in an earlier study and
extended in this paper, sets a limit of 2500K for the $T_{\rm eff}$ of an
accretion disk that may be present in the low state. This limit is in
conflict with two recent models of the VY Scl phenomenon. 

\end{abstract}



\section{Introduction}

Cataclysmic variables (CVs) are binary stars consisting of a white dwarf (WD) and
a late spectral type secondary which fills, or almost fills, its Roche lobe
and transfers mass to the WD. CVs divide into a number of
subclasses \citep{w95} and the object of our study,
MV Lyrae, is a novalike CV of the VY Scl subclass.
A defining characteristic of this subclass is a non-periodic alternation
between high and low luminosity states, differentiated by a change in mean brightness
of more than one magnitude between states. 
This alternation usually
takes place on intervals of hundreds of days.
 
The orbital period of MV Lyr, $3.19^{\rm h}$, places it close to the long-period end
of the period gap \citep{w95}. Its high-low range is 12.2-18, inferred from Fig.4 of  \citet*{hk04}.
While it is widely argued that evolution of
CVs with periods longer than the period gap is controlled by magnetic 
braking \citep{p1984,k1988}, and evolution shortward of the gap is driven by
gravitational radiation \citep{k1988}, this scenario has been recently challenged 
by, e.g., \citet*{aps03}, \citet*{ks02}, and \citet*{sk02}. 
Patterson (1984) developed an empirical mass transfer relationship for CVs which
predicts a mass transfer rate of
$2.5{\times}10^{-10}{\cal M}_{\odot}{\rm yr}^{-1}$ at the orbital period of MV Lyr.
A theory of magnetic braking \citep{h1988} shows that the mass transfer rate from
the secondary is a decreasing function with decreasing orbital period, reaching a value of
$1.0{\times}10^{-9}{\cal M}_{\odot}{\rm yr}^{-1}$ at the top of the period gap, where MV Lyr
is found,
for the parameters used in their study.	Some authors, as discussed below, adopt even larger
mass transfer rates at the top of the period gap.

If the magnetic braking mechanism that drives evolution of the secondary star
is suddenly turned off at an orbital period of $3^{\rm h}$, then the secondary
approaches thermal equilibrium on a Kelvin-Helmholtz timescale. Consequently,
mass transfer continues at a decreasing rate for ${\approx}10^4{\rm yr}$
\citep{r1988}. But the alternation between high and low states in VY Scl stars is
on a much shorter timescale and requires a separate mechanism to switch the mass
transfer on and off. \citet*{lp94} discuss the problem and
propose starspots on the secondary star, which pass across the L1 point
and cause reduced mass transfer. This proposal has been further
developed \citep{kc98} with application of a time-dependent code to model the
variable accretion disk. (See also \citealt*{sgc2000}). But the \citet{kc98} model produced 
the unwanted result of dwarf-nova
type outbursts in the low state, after mass transfer had been switched off.
A suggested escape was maintenance of the inner accretion disk in a permanent
high state through irradiation by the WD 
(Leach et al. 1999, hereafter L1999). L1999 concluded that the 
presence of a
40,000K WD does suppress dwarf nova outbursts.	The L1999 study adopted a 
high state mass transfer
rate of $1.1{\times}10^{-8}{\cal M}_{\odot}{\rm yr}^{-1}$
that was chosen to guarantee that hydrogen is fully ionized
at the outer accretion disk boundary.
According to L1999, cessation of mass transfer produces a rapid transition to a
state of low but non-zero viscosity. The accretion disk remains nearly intact
during the low state, transferring only a few percent of the accretion disk mass to 
the WD. 
Because of non-zero viscosity, the accretion disk still radiates at a low level
during the low state.
Hameury \& Lasota  
(2002, hereafter HL02) show that the adopted WD mass in L1999 (0.4${\cal M}_{\odot}$)
is important to the suppression of outbursts, and that a WD mass of
0.7${\cal M}_{\odot}$ would have produced outbursts. HL02 alternatively propose that VY Scl
stars have magnetic WDs and that the magnetic field truncates the accretion
disk. HL02 show that, for a WD mass of 0.7${\cal M}_{\odot}$
and a magnetic moment of ${\mu}=5{\times}10^{30}{\rm Gcm^3}$, no outbursts occur
and the change in the $V$ magnitude, from high state to low state, is 5 magnitudes. 
 
Hoard et al. (2004, hereafter H2004) used the BINSYN suite \citep{lin96} to
calculate a system model and corresponding synthetic spectrum for MV Lyr in a 
recent low state.
The synthetic spectrum accurately fits {\it FUSE}
spectra and contemporaneous optical spectra, as well as {\it IUE} spectra from
a prior low state.
The H2004 discussion considers the 
connection of the low state study
to earlier investigations of MV Lyr and 
shows that the low state
can be understood in terms of a naked hot WD with a temperature
of 47,000K, a photosphere log $g$ of 8.25, and a metallicity of 0.3 solar.
The log $g$, together with the mass-radius relation for zero-temperature carbon WD models
\citep*{hs61} selects a model with ${\cal M}_{\rm wd}=0.73{\cal M}_{\odot}$, estimated
error of $0.1{\cal M}_{\odot}$, and radius of $0.01067R_{\odot}$.
The secondary star
fills its Roche lobe and is cooler than 3500K; it contributes
nothing to the FUV flux and a varying amount to the optical flux, from 10\% at
5200\AA~to 60\% at 7800\AA. If an accretion disk is present it contributes negligibly
to the observed spectra. Assuming no disk is present, the model optical light curve,
for a system orbital inclination of $12{\degr}$ taken from the literature, has an amplitude
approximately 50\% larger that of the observed optical light curve. The scaling of the
model spectrum to the observed data leads to a distance of $d=505{\pm}50$pc to MV Lyr. 

In \S2 of this paper we describe the program suite used to analyze the observed spectra.
In \S3 we further investigate the low state and derive improved
system parameters. In \S4 we show that it is possible to augment the low state
system with an accretion disk
with artificially elevated $T_{\rm eff}$ at large radii and achieve a close
fit to intermediate state {\it IUE} archival spectra. In \S5 we show that our models
produce a predicted ${\Delta}V$, low state to intermediate state, that is one
magnitude larger than observed, indicating a problem with the models. In \S6 we
show that an isothermal 14,000K accretion disk extending half way to the tidal cutoff 
radius resolves the ${\Delta}V$ problem and provides a fair fit to the {\it IUE}
spectra.
\S7 presents a model fit to a new high state HST spectrum, \S8 considers
the predicted luminosity change low state to high state, a general discussion is in \S9
and \S10 summarizes our results.

\section{Description of the BINSYN program suite}

The BINSYN suite \citep*{lin96}, used in the analysis presented in this paper, is a 
set of programs for calculating synthetic spectra and
synthetic light curves of binary star systems with or without accretion disks and in
circular or eccentric orbits. 
The model stars have photospheres of assigned Roche potentials, defined by grids
of points with associated photospheric segments. 
Program modules assign $T_{\rm eff}$
values to individual stellar points based on adopted polar $T_{\rm eff}$ values, bolometric albedos,
and gravity darkening exponents. Model stars may rotate synchronously or non-synchronously
up to critical rotation.
Program modules calculate log $g$ values at stellar photosphere
points based on adopted Roche potentials and stellar masses, while other modules determine whether 
a given point is
visible to the observer, for a point on a star, the accretion disk face or the accretion
disk rim, and for the current values of orbital inclination and orbital longitude.
Calculation of a synthetic spectrum uses a set of precalculated source synthetic spectra for
each system object (stars, accretion disk face, accretion disk rim). The source spectra for 
a given object typically form a two-dimensional
set in $T_{\rm eff}$ and log $g$ that encompasses the extreme $T_{\rm eff}$ and log $g$
values on the photosphere of the given object. Program modules interpolate an
individual synthetic spectrum to each stellar photospheric point and integrate over that star, with
allowance for point visibility; the routines properly allow for Doppler shift and line broadening.

An accretion disk is specified in BINSYN by an inner and outer radius, and the accretion disk
is divided into a specified number of cylindrical annuli, with each annulus divided
into a specified number of segments. 
Radiation characteristics of the accretion disk depend on a temperature profile, $T(R)$, of the
accretion disk annuli.
We reserve the term "Standard Model" to designate a $T(R)$ relation

\begin{equation} 
T(R) = W\{\frac{3G{\cal M}\dot{\cal M}}{8{\pi}{R_*}^3{\sigma}}[1-{(\frac{R_{*}}{R}})^{1/2}]\}^{1/4}
(\frac{R}{R_*})^{\beta},
\end{equation}
with $\beta=-0.75$ for the Standard Model
\citep[hereafter FKR]{fkr95}. $W$ is a normalizing factor, (not a dilution factor); 
$W =1.0$ for
the Standard Model.	
For default conditions, our definition is algebraically identical
to the FKR definition.

The accretion disk is flared, in the sense that
successive annuli step up in height between the inner radius and a specified
semi-height at the outer radius.
Let $H$ be the semi-thickness at the outer radius $r_a$, and let $r_b$ be the inner radius.
Then the semi-thickness at radius $r$, $h(r)$, is $h(r)/H = (r-r_b)/(r_a-r_b)$. See Table~1. 
In calculating a model for an accretion disk system, BINSYN, by default, assigns
$T_{\rm eff}$ values to annuli based on the Standard Model, appropriate to an
assumed mass transfer rate. An option permits modification of the assigned annulus
$T_{\rm eff}$ values to include calculated irradiation by the WD, based on a bolometric 
albedo formalism.
Calculation of a synthetic
spectrum for the accretion disk, as part of the calculation of the system synthetic
spectrum, involves interpolation among previously calculated source spectra	(with their
individual $T_{\rm eff}$ values)
to produce a synthetic spectrum for each annulus, for its $T_{\rm eff}$ as assigned by
BINSYN. The accretion disk source spectra may consist largely of synthetic spectra for
annulus models produced by TLUSTY \citep{h88,hl95}. 
This calculation is followed by 
integration over all
annuli, with detailed allowance for Doppler shifts and sources of line broadening,
and with proper allowance for visibility to the observer. The
number of source spectra typically will differ from the specified number of annuli.
The programs finally output the calculated flux as function of wavelength for the individual 
objects as well
as their sum, which represents the system synthetic spectrum at the current orbital longitude.
A parallel procedure produces synthetic light values, repeated for
a grid of orbital longitudes and so leads to light curves for specified wavelengths. 
Additional details are in \citet*{lin96}.

\section{Fit to the low state optical spectra}

\begin{figure}
\figurenum{1}
\plotone{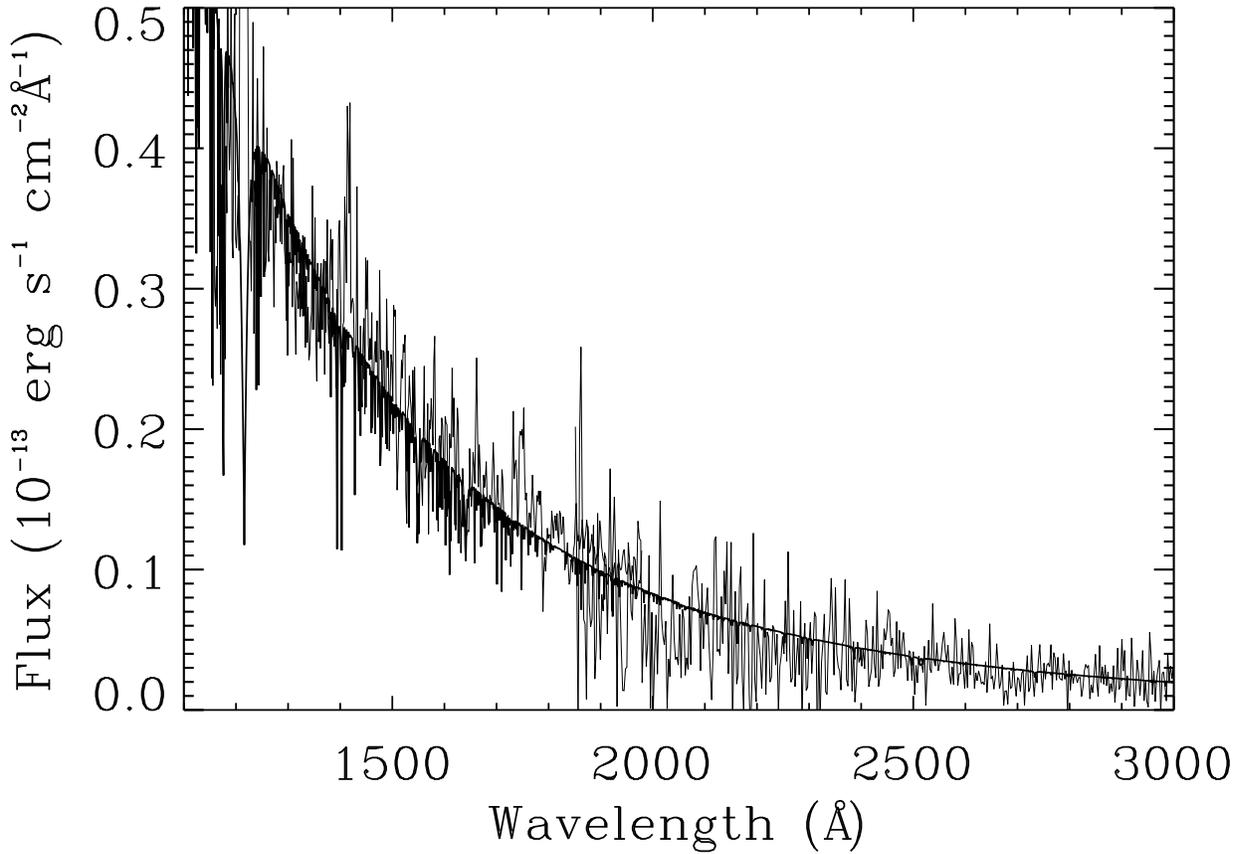}
\vspace{0.1cm}
\caption{Fit of synthetic WD spectrum to low state {\it IUE} spectrum of MV Lyr.
The light line is the {\it IUE} spectrum, the heavy line (colored blue in the
electronic edition) is the
synthetic spectrum. 
{\it (See the electronic edition of the Journal for a color	version of this figure.)}
\label{fig1}}
\end{figure}

In fitting the optical spectra, H2004 found two difficulties: (1) The polar
$T_{\rm eff}$ of the secondary was cooler than 3500K, which is the temperature of 
the coolest (Kurucz) model
atmosphere available to us. (2) The optical spectra were observed through
thin clouds so an empirical scaling factor was applied to the observed fluxes to
match them to the synthetic spectrum (the latter was scaled to fit the FUV spectrum).

Fig.~8 of H2004 shows the model fit to observed spectra, covering the interval 
930\AA~to 7800\AA,
including archived {\it IUE} spectra SWP10905, covering 1150\AA~to 1978\AA, and LWR09590,
covering 1851\AA~to 3349\AA, which were obtained 
on 1980 Dec. 27. 
Fig.~1 of this paper presents the model
fit to the same low state {\it IUE} spectra in much greater detail than in H2004.
Note that Fig.~8 of H2004 is a logaritmic plot while Fig.~1 of this paper plots flux
directly.
In producing Fig.~1, 
the spectra were divided by the same scaling factors
applied in the H2004 plots	
($2.43{\times}10^{29}$ for the
WD synthetic spectrum, $1.0{\times}10^{-13}$ for the {\it IUE} spectra),
set by the fit to the FUSE spectrum.
The mean residual of the fit to the 877 points between 1150\AA~and 3000\AA~was 0.012 and
the mean absolute residual was 0.038. These residuals apply to the scaled spectra.
An important conclusion from the Fig.~1 fit is that there is no evidence for the 
presence of an accretion disk. In a test of the sensitivity of the fit to added 
flux from a putative accretion disk, we find empirically that
addition of 0.005 
flux units (Fig.~1 ordinates)
produces a visually detectable displacement
of the summed spectrum from the Fig.~1 fit. 
At 3000\AA, the secondary star makes a calculated contribution of 0.1\% to the system
flux.

H2004 fit an optical (unfiltered) light curve of MV Lyr (their Fig.~7).
This fit was derived from the BINSYN spectral model described above. It assumes an
orbital inclination for the CV of $i = 12{\arcdeg}$ from
\citet*[hereafter SPT95]{spt95}. However, the corresponding
amplitude of the model light curve is 50\% too large. The orbital period light variation
results from the variable presentation of the irradiated secondary component to
the observer. 
As the assumed inclination decreases from
$i = 12{\arcdeg}$, the light amplitude decreases monotonically.
An improved fit \citep{l2005} implies a system inclination of $i = 7{\pm}1{\arcdeg}$,
which we adopt for this analysis.

A remark is in order on the absolute accuracy of the results obtained in this paper.
A key step is in the determination of the WD mass and radius in units of the solar values by 
comparison between the MV Lyr WD log $g$
from our analysis of the FUSE spectrum and the log $g$ values for the \citet*{hs61}
carbon WD models. The mass ratio, $q$, comes from relatively uncertain radial velocity
measurements (SPT95), and this value determines the mass of the secondary star
and, since the secondary fills its Roche lobe, the size of the secondary star.
The fit of our synthetic spectra to FUV data, in all cases, applies a scaling
divisor, $2.43{\times}10^{29}$, to observed spectra which were divided by $1.0{\times}10^{-13}$
(see the ordinate labels).
The scaling divisor may be inaccurate by 10\% to 15\%.

We present here an improved representation of the secondary star,
using the NextGen synthetic spectra for M stars, version 5 
\citep{haus99}. We find a good fit to the observed spectra
with a polar $T_{\rm eff} = 2600$K. Irradiation by the 47,000K WD produces a calculated 
L1 point 
$T_{\rm eff} = 5344$K on the secondary star. BINSYN interpolated among seven NextGen
synthetic spectra with $T_{\rm eff}$ values between 2600K and 5400K to determine
flux values for individual secondary star photospheric segments.
The fit of the system synthetic spectrum to the optical spectra is shown in Fig.~2. 
The optical spectra	were multiplied by an empirically determined factor of 2 to 
correct for obscuration by clouds.

Recapitulating, we have developed a model for the low state of MV Lyr consisting of a 47,000K WD
and a secondary star with a polar $T_{\rm eff}$ of 2600K. The secondary star fills its Roche lobe
and is irradiated by the WD. The model has no accretion disk, and differs from the model in H2004
only in the improved representation of the secondary star in the calculated synthetic spectra.
This model produces an excellent fit to the FUSE data, the low state {\it IUE} data, and the optical
data.
We proceed to compare our model to the low state predictions of L1999 and HL02.

\begin{figure}
\figurenum{2}
\plotone{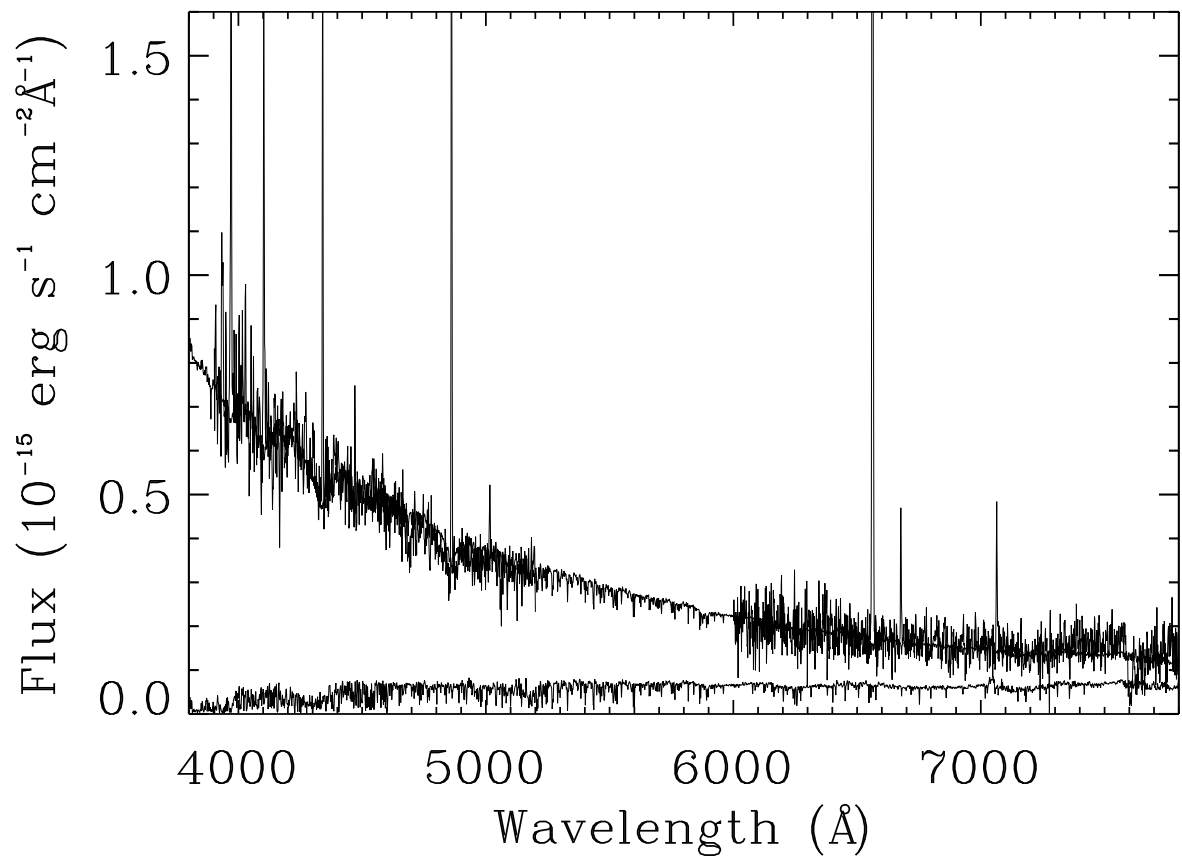}
\vspace{0.1cm}
\caption{Fit of synthetic spectrum to low state optical spectra of MV Lyr.
The light line indicates the optical spectra, the heavy line (colored blue in 
electronic edition) is the system
synthetic spectrum. The secondary star contribution to the system synthetic
spectrum is the lower spectrum in the plot. 
{\it (See the electronic edition of the Journal for a color	version of this figure.)}
\label{fig2}}
\end{figure}

\begin{figure}
\figurenum{3}
\plotone{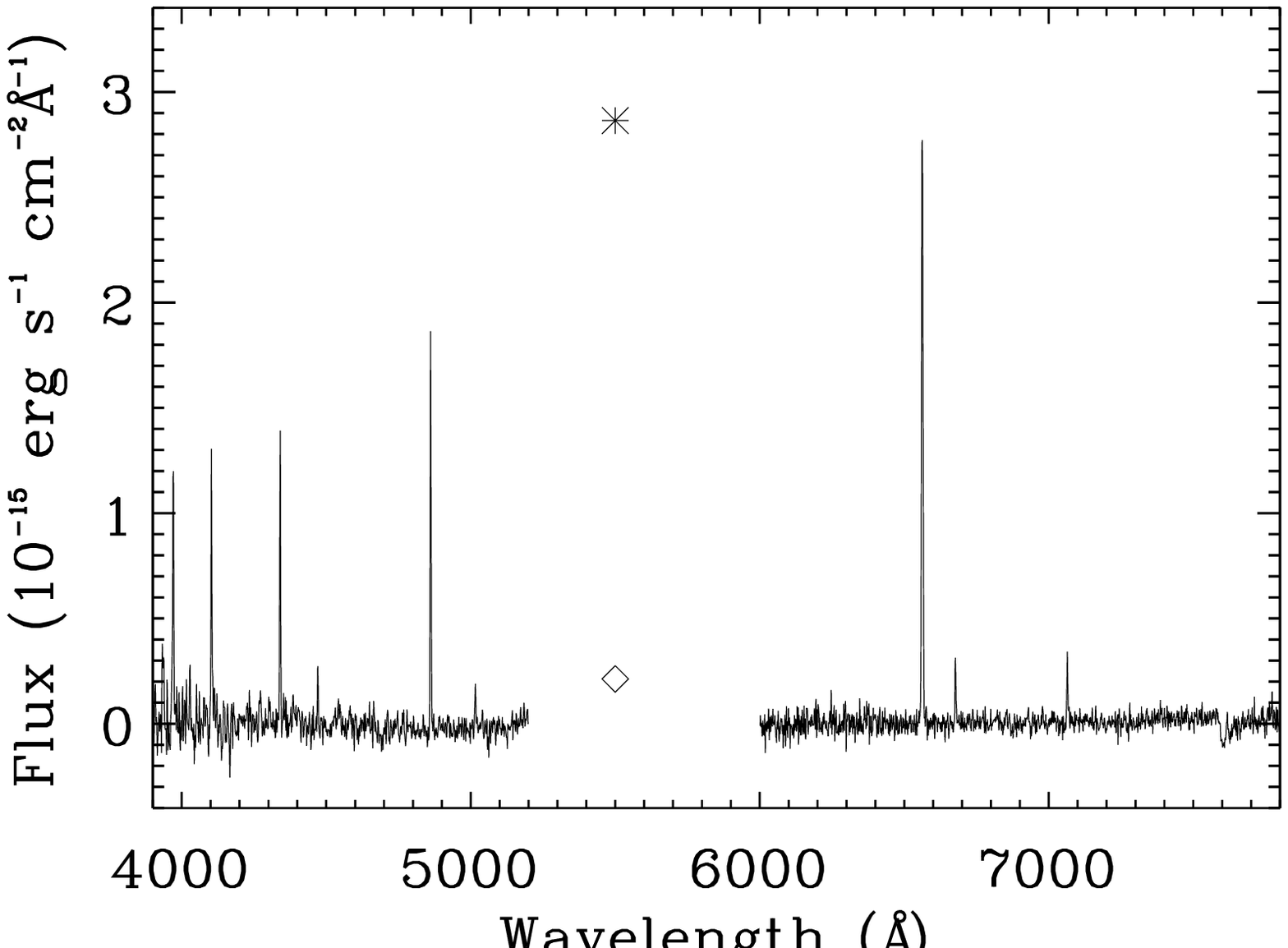}
\vspace{0.1cm}
\caption{The O(optical spectra)-C(synthetic spectrum) residual spectrum
from Fig.~2. The asterisk
marks the 5500\AA~flux of the L1999 model (see text) if it is placed at the
distance of MV Lyr, while the diamond marks the corresponding datum for the HL02
model (see text).
The feature at 7600\AA~is due to the molecular oxygen "A"
absorption band in the terrestrial atmosphere. The emission lines are produced
on the secondary star by irradiation from the primary star.
\label{fig3}}
\end{figure}

L1999 (their Fig.~3) predict an accretion disk luminosity in the low state of
${\rm F}_{5500}=7.0{\times}10^{27}\,\mathrm{erg\,s^{-1}\AA^{-1}}$. 
Fig.~3 plots the residual of the L1999
accretion disk (asterisk), for the 505pc (H2004) distance of MV Lyr.
The residual is ${\rm F}_{5500}/d^2$, where $d$ = 505pc. 
The L1999 accretion disk flux at
5500\AA~would have to be approximately 60 times fainter to accord with the residual plot.

We calculated a synthetic spectrum for the example in HL02, with ${\mu}_{30}=5$,
with ${\cal M}_{\rm wd}=0.73{\times}{\cal M}_{\odot}$
and mass ratio $q=0.43$. 
The calculated inner radius
for the HL02 example (equation 4) is $r_{\rm in}=5.7{\times}10^8$ cm and the MV Lyr WD
radius is $r_{\rm wd}=7.9{\times}10^8$ cm. Consequently we set the inner accretion disk 
radius equal to the WD radius, the outer radius equal to the tidal truncation radius,
and the mass transfer rate equal to ${\dot{\cal M}}=2{\times}10^{17}{\rm gm~{s^{-1}}}$ (HL02 value).
We used the synthetic spectrum of the calculated accretion disk alone in the following comparison.
We divided the output flux in the synthetic spectrum by 100.0,
the factor by which the HL02 model luminosity drops between the high and low states,
to provide the HL02 model flux to compare with our low state data.
As with L1999, this calculated low state luminosity is a residual, 
[(system model)-(BINSYN low state model)], shown in
Fig.~3 by a diamond.
It could be argued that division by 100.0 for all wavelengths is too simplistic in 
transforming the HL02 model to an equivalent low state, particularly since we have no observational 
data in the 5000\AA~to 6000\AA~interval. Our Fig.~2 shows that the low state synthetic spectrum 
varies smoothly over this wavelength interval, as does our intermediate state model and high state 
model, discussed in the following sections (not shown in separate figures). Thus, we believe
the division by 100.0 is appropriate.
The HL02 flux needs to be smaller by approximately a factor 5-10 to agree
with the spectral residuals. Thus, we find a discrepancy between the predictions of both
L1999 and HL02 for the MV Lyr low state and our fit to available low state data.

How cool would an MV Lyr accretion disk have to be, in the low state, to
avoid detection in optical spectra? 
Tests with an isothermal accretion disk, represented by a black body and truncated at 
$r=1.7{\times}r_{\rm wd}$ (see the following text for a discussion of the truncation radius), 
demonstrate
that an accretion disk must have a $T_{\rm eff}$ less than 2500K to avoid a conflict
with our optical spectra (a 2500K accretion disk would provide a detectable 3\% of the
system flux at 7800\AA).
A study of TT Ari \citep{g1999} obtained comparable results. 
A possible accretion disk would need to be truncated at $r=12r_{\rm wd}$ and 
have a
$T_{\rm eff}$ less than 3000K to satisfy the low state optical 
spectra.

\section{Fit to the intermediate state {\it IUE} spectra}

The {\it IUE} archives include 10 spectra of MV Lyr. None of these were obtained
during a high state, and two, SWP07296 and
LWR06288, were obtained on Dec. 2, 1979 during an intermediate state.
Intermediate states are roughly one magnitude fainter than high states.
  
Fig.~8 of
H2004 shows the intermediate state {\it IUE} spectra on the same plot as the fit to the 
FUSE/{\it IUE}/optical low state
spectra, for comparison. A striking feature of the intermediate state {\it IUE} spectra 
is their
much lower (i.e., less blue) spectral gradient than the low state {\it IUE}
spectra.
This is surely due to the presence of an optically thick accretion disk.

We calculated a series of 
annulus
models, using TLUSTY version 200 \citep{h88,hl95} for individual mass transfer rates of
$1.0{\times}10^{-9}{\cal M}_{\odot}{\rm yr}^{-1}$,
$4.0{\times}10^{-9}{\cal M}_{\odot}{\rm yr}^{-1}$, and
$1.0{\times}10^{-8}{\cal M}_{\odot}{\rm yr}^{-1}$. These latter models were spaced in values of
$r/r_{\rm wd}$ between 2.5 and an outer value of $r/r_{\rm wd}=28.5$, $T_{\rm eff}$=10,620K.
TLUSTY includes a large number of optional control flags that control the program performance.
We preserved the default control flags, and thereby produced LTE thin disk models with thin
photospheres.
Convergence 
problems intervened for	larger values of $r/r_{\rm wd}$.
We used a viscosity parameter ${\alpha}=0.1$ \citep{SS73} for all models.
The models included opacity contributions from the
30 most abundant periodic table elements. We used these models to calculate source
(see above) synthetic spectra, via SYNSPEC 
version 48
\citep{hsh85}, for all models, with
the same opacities used for the model calculations. 
We included synthetic spectra for Kurucz stellar model atmospheres, calculated with SYNSPEC,
to handle assigned temperatures
cooler than the 10,897K limit described above. Comparison among the annulus source synthetic 
spectra
for the different tabular mass transfer rates showed that spectra for different mass transfer 
rates but similar $T_{\rm eff}$ values are essentially identical. Consequently, it is
permissible to fix the
source synthetic spectra of the annuli at some specific tabular mass transfer rate, 
but change the
BINSYN mass transfer rate for a new simulation, and still produce a valid system synthetic 
spectrum. 

The accuracy of the accretion disk synthetic spectrum is sensitive
to the $T_{\rm eff}$ spacing of the source synthetic spectra. Linear interpolation
between source spectra differing in $T_{\rm eff}$ values by several thousand Kelvins
can produce an inaccurate interpolated spectrum. 
After extensive experiment, we adopted
a "universal" set of 69 source spectra, of which 62 were TLUSTY models. These latter models
had $T_{\rm eff}$ values spaced at roughly 500K intervals between 10,620K and 31,564K and
a slightly larger step between 31,564K and 63,735K.
The remaining synthetic spectra were for Kurucz models covering the $T_{\rm eff}$ range
5000K--10,000K. In the wavelength range of interest, contributions from ${\approx}7000K$ or
cooler $T_{\rm eff}$ values are very small or negligible.

Our initial study of the intermediate state adopted a variety of assumed mass transfer rates, 
together with
a Standard Model accretion disk.
We used the system parameters of Table~1, and BINSYN in a diagnostic
mode, to attempt a fit to the intermediate state {\it IUE} spectra.	With the exception of 
the parameters $D$, $i$, $T_{\rm eff,s}$, $r_a$, $r_b$, and $H$ (see Table~1 comments),
the Table~1 parameters
for the WD and secondary star are the same as in H2004.
Our model divides the accretion disk
into 66 annuli, with each annulus divided into 90 azimuthal segments. 
The system synthetic spectrum
is the integrated sum of the contributions of the WD, secondary component, accretion
disk, and accretion disk rim.
(Note that, at an inclination of 
$7{\arcdeg}$, the accretion disk rim contributes a negligible observable effect.)

\begin{deluxetable}{llll}
\tablewidth{0pt}
\tablecaption{BINSYN Model System Parameters}
\tablehead{
\colhead{parameter} & \colhead{value} & \colhead{parameter} & \colhead{value}}
\startdata
${ M}_{wd}({ M}_{\odot})$  &  0.73	 & $r_{wd}$      &   0.01017\\
{\it q}                     &  0.43	     & log $g_{wd}$  &   8.25\\
P(d)                        &  0.1329	 & $r_s$(pole) &  0.28782\\
$D(R_{\odot})$              &  1.11120   & $r_s$(point) & 0.40362\\
${\Omega}_{wd}$                &  98.74     & $r_s$(side)  & 0.30011\\
${\Omega}_s$                &  2.74      & $r_s$back)   & 0.33258\\
{\it i}{\degr}              &   7.0		 & log $g_s$(pole) & 4.95\\
$T_{\rm eff,wd}$(K)         &  47,000       & log $g_s$(point) & 3.61\\
$T_{\rm eff,s}$(K,pole)         &  2600       & log $g_s$(side)  & 4.87\\
$A_{wd}$                       &  1.0       & log $g_s$(back)  & 4.68\\
$A_s$                       &  0.5       & $r_a(R_{\odot})$ & 0.47\\
$b_{wd}$                       &  0.25      & $r_b(R_{\odot})$ & 0.01\\
$b_s$                       &  0.08      & $H(R_{\odot})$    & 0.0110\\
\enddata
\tablecomments{$wd$ refers to the WD; $s$ refers to the secondary star.
$D$ is the component separation of centers,
${\Omega}$ is a Roche potential. Temperatures are polar values, 
$A$ values are bolometric albedos, and $b$ values are 
gravity-darkening exponents. $r_{wd}$ is in units of component 
separation, not solar radii. $r_a$ specifies the outer radius 
of the accretion disk, set at the tidal cut-off radius, 
and $r_b$ is the accretion disk inner radius, while $H$ is 
the semi-height of the accretion disk rim.}  
\end{deluxetable}

We find that it is not possible to fit the intermediate state
{\it IUE} spectra by adding a Standard Model accretion disk to 
the low state model of H2004 and using our TLUSTY annulus synthetic spectra. 
The radial temperature gradient in the Standard Model
produces either too much flux at short wavelengths or too little flux at long
wavelengths. 
In making this (and all subsequent) comparisons with the
{\it IUE} spectra we maintained the same scaling divisors, for both the {\it IUE}
spectra and the synthetic spectrum, that were adopted for the low state comparison.
Based on the measured $E(B-V) = 0.0$ \citep{sd82} we applied no 
interstellar
extinction corrections to the {\it IUE} spectra extracted from the archive.
Fig.~4 illustrates the discrepancy produced by adding an accretion disk component 
with a mass transfer rate of
$1.0{\times}10^{-9}{\cal M}_{\odot}{\rm yr}^{-1}$. Note that this rate equals
the value given by Hameury et al. (1988) at the top of the period gap.
This mass transfer rate, and a Standard Model disk, produces
disk $T_{\rm eff}$ values ranging from 23,765K at the inner disk edge to 4222K at the
outer disk edge, with a temperature maximum of 34,972K at $r=1.3611{\times}r_{\rm wd}$.
The absolute flux from the accretion disk, at the Earth, is a strong function of the 
mass transfer rate and the approximate fit in Fig.~4 indicates that the adopted
mass transfer rate is nearly correct, i.e., the {\it mean} residual over the wavelength
interval plotted is small. The synthetic spectrum is too blue. A larger mass transfer
rate increases the flux at all wavelengths, making the mean residual larger, and produces
a synthetic spectrum that is too blue. A smaller mass transfer rate decreases the flux at 
all wavelengths, again making the mean residual larger, and produces a synthetic spectrum
that is too blue.

\begin{figure}
\figurenum{4}
\plotone{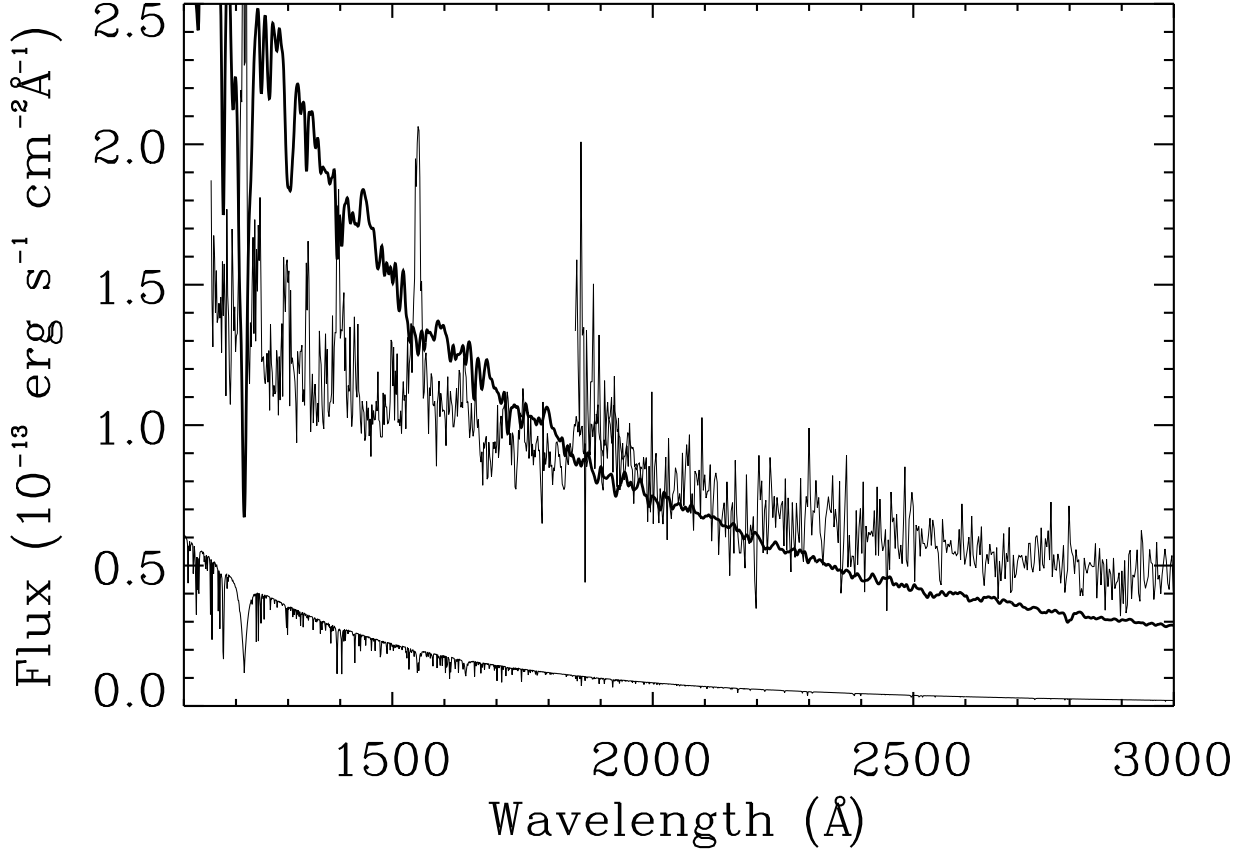}
\vspace{0.1cm}
\caption{Fit of synthetic spectrum to intermediate state {\it IUE} spectra of MV Lyr.
The light line is the {\it IUE} spectrum; the heavy line (colored blue in electronic edition)
is the 
synthetic spectrum. The synthetic spectrum includes a Standard Model accretion
disk with a mass transfer rate of
$1.0{\times}10^{-9}{\cal M}_{\odot}{\rm yr}^{-1}$. The synthetic spectrum was convolved
with a Gaussian 5\AA~FWHM broadening function to match the {\it IUE} spectral resolution.
The low state synthetic spectrum from Fig.~1 (lower spectrum) is shown for comparison. 
{\it (See the electronic edition of the Journal for a color	version of this figure.)}
\label{fig4}}
\end{figure}

The Fig.~4 residuals suggest a departure from a Standard Model.
Stationary models
are possible \citep{l2001} whose $T(R)$ profiles differ from the Standard Model. Examples
include systems in which irradiation by the WD is important, or systems in which the mass 
transfer stream directly heats the outer regions of the accretion disk \citep{vd89,vd92,sm94},
(but see \citealt{al98}), or
there is an increased mass flux through the outer part of the accretion disk for
unspecified reasons \citep*{kh86}.
In all of our synthetic spectra, the synthetic spectrum has absorption
lines where the {\it IUE} spectrum shows emission lines. 
SPT95 argue that the broad emission lines in
the "high" ({\it V}${\sim}13-14$) state arise from the accretion disk, in contrast to 
narrow (Fig.~2) emission lines
produced on the irradiated secondary when MV Lyr is in a low state.

Two variations of the Standard Model T(R) relation affect the corresponding
system synthetic spectrum: (1)
a change in the functional dependence of $T(R)$ on $R$ \citep{ow03}, and (2) 
truncation of the accretion disk at some inner radius larger than the WD radius \citep{L94}.
Physically, truncation can occur either by evaporation of the inner disk annuli \citep{h2000}
or by accretion disk interaction with a magnetic field associated with the WD (HL02).

\begin{figure}
\figurenum{5}
\plotone{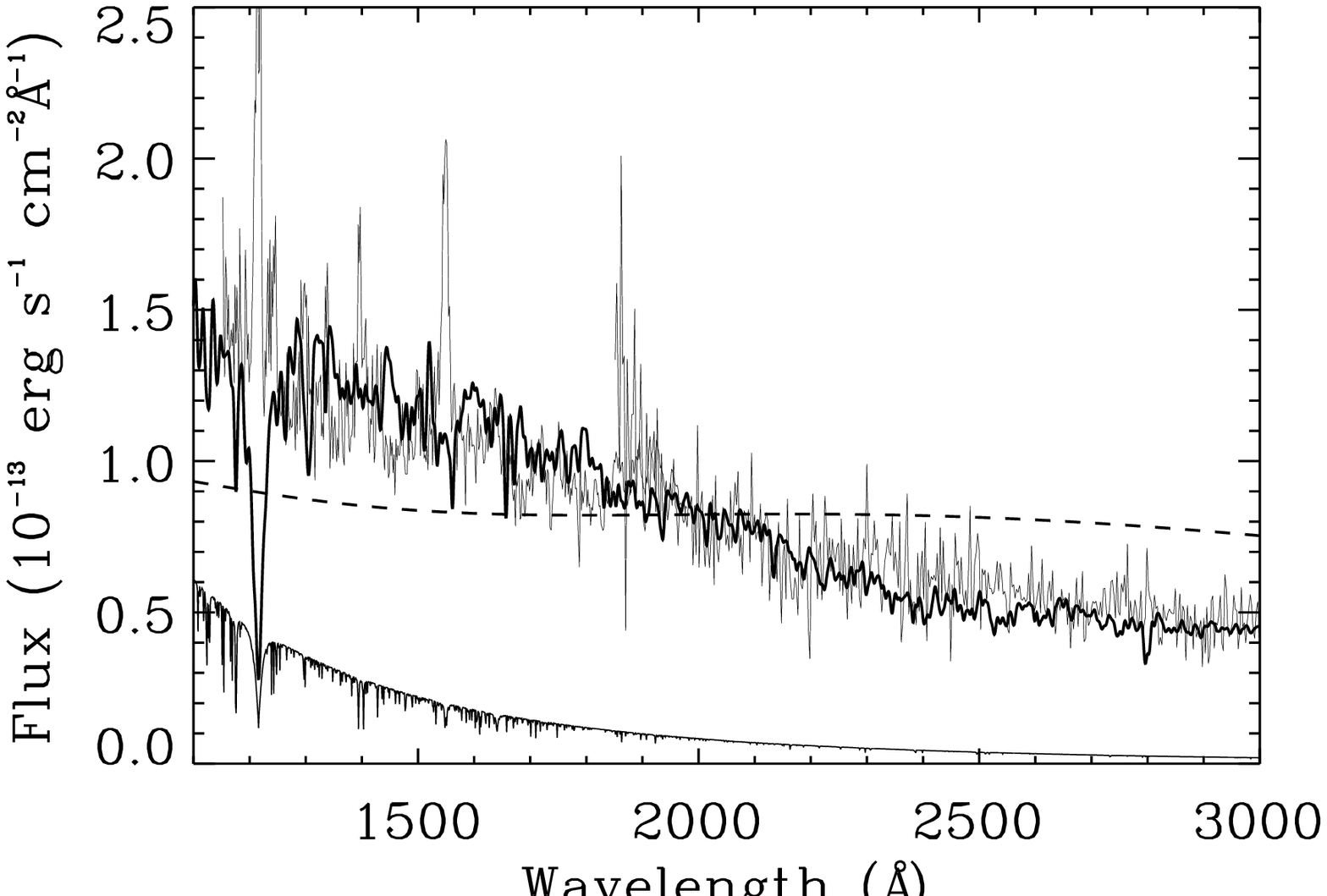}
\vspace{0.1cm}
\caption{As in Fig.~4 but with an accretion disk truncated at $r=1.7{\times}r_{\rm wd}$,
a $T(R)$ corresponding to a mass transfer rate of 
$6.0{\times}10^{-10}{\cal M}_{\odot}{\rm yr}^{-1}$, and
an isothermal 9500K accretion disk beyond $r=11.5{\times}r_{\rm wd}$.
The dash line represents a corresponding black body model.
The light line is the {\it IUE} spectrum; the heavy line (colored blue in electronic edition)
is the synthetic spectrum. The low state synthetic spectrum from Fig.~1 (lower spectrum)
is shown for comparison.
{\it (See the electronic edition of the Journal for a color	version of this figure.)}
\label{fig5}}
\end{figure}

We have investigated both options. The first permits a good fit to the {\it IUE} spectra
but the models are slightly inferior to the second option models.
The second option
truncates the accretion disk at an inner radius larger than the WD radius \citep{L94} 
and preserves the Standard Model $T(R)$ functional dependence on $R$. 
Since
the distance to MV Lyr is a fixed quantity, as the truncation radius of 
the accretion disk is increased, producing an initial reduction in flux from the model,
and a resulting discrepancy with observed spectra, a compensating increase in the
adopted mass transfer rate is necessary to restore agreement with the observed
spectra. 
We started with the mass transfer rate of Fig.~4, then incrementally increased the
truncation radius and made compensating increases in the mass transfer rate. The
quality of the successive
fits improved but remained perceptably discrepant (too blue) even up to a limiting
truncation radius of $r_b/r_{\rm wd}=18$. At this 
truncation radius, 50\% of the tidal cutoff radius, the
necessary mass transfer rate becomes $1.0{\times}10^{-8}{\cal M}_{\odot}{\rm yr}^{-1}$,
close to the one used in L1999, and one dex larger than the
value given by Hameury et al. (1988) at the long period boundary of the period gap.	
We conclude that a Standard Model T(R) with
a truncated accretion disk model produces a synthetic spectrum that is too blue and so 
does not fit 
the observed
MV Lyr {\it IUE} spectra.

Our initial models
maintained the WD $T_{\rm eff}$ of 47,000K (H2004). \citet{si95} has shown
that mass transfer can cause compressional heating of WDs in CV systems by
10,000K-20,000K for mass transfer rates of order 
$1.0{\times}10^{-8}{\cal M}_{\odot}{\rm yr}^{-1}$ or larger
(also see \citealp*{tb02}). 
We tested the model sensitivity to the WD $T_{\rm eff}$	by
substituting a NLTE, $T_{\rm eff}$
= 52,000K, log $g$=8.25 WD synthetic
spectrum for the 47,000K synthetic spectrum. The change in the system synthetic spectrum,
for the wavelength range 1100\AA-3000\AA~was almost undetectable at 3000\AA. 
In view of our much smaller mass transfer rate than those studied by Sion, 
we did not test larger $T_{\rm eff}$ values and we do not consider this topic further.

Our extensive experiments ultimately led us to a best-fitting model with a
mass transfer rate of $6.0{\times}10^{-10}{\cal M}_{\odot}{\rm yr}^{-1}$, a truncation
radius of $r_b=1.7{\times}r_{\rm wd}$,  
and with $T(R)$, which initially follows the Standard Model, set to 9500K for radii 
larger than $11.5{\times}r_{\rm wd}$. The $T_{\rm eff}$ of the innermost annulus, including
the correction for irradiation by the WD, is 30,195K. 
Fig.~5 shows the fit of this solution
to the {\it IUE} data.
The mean residual of the fit to the 877 points between 1150\AA~and 3000\AA~was 0.013 and
the mean absolute residual was 0.158. These residuals apply to the scaled spectra. 
The synthetic spectrum is too blue.
The dash line represents a corresponding black body model in which all source synthetic spectra
have been replaced by black body spectra of the same $T_{\rm eff}$. The black body model is not 
blue enough.

\citet*{kh86} (their Fig.~2) provides optical thickness as a function of wavelength
for two TLUSTY-based accretion disk models. A rough interpolation indicates that our models 
likely are optically thick. 
If the accretion disk remains
nearly Keplerian, our Fig.~5 model, for equal bolometric luminosity, would correspond to a 
Standard Model of mass
transfer rate higher than $6.0{\times}10^{-10}{\cal M}_{\odot}{\rm yr}^{-1}$, and likely
near $1.0{\times}10^{-9}{\cal M}_{\odot}{\rm yr}^{-1}$.

We have not attempted to calculate irradiated models of the	individual accretion disk annuli,
a possible option in TLUSTY. Because of the disk truncation, no boundary layer has been 
included in the synthetic spectrum.

\section{Predicted change in $V$ magnitude between the low and intermediate states}

The AAVSO light curve shown in H2004, and the Roboscope curve shown in \citet{hk04},
covering 1990--2004, reveals that MV Lyr 
exhibits high states lasting for
2--6 months, followed by low states lasting for similar lengths of time.
Typical high state $V$ values are 12.2-12.8, while low state $V$ values are in
the range 16-18.
Intermediate state
magnitudes show considerable variation, and in the $V$ range 13--15.
Earlier $B$ and $V$ light curves (\citealp[hereafter RRM]{r1993}) cover the interval 1968 to 1991
and show that MV Lyr was in an extended low state from 1979 to 1989, with occasional
transitions to intermediate states followed by returns to the low state. 
 
RRM tabulate $B = 14.6$ on Oct.17, 1979, $B = 13.9$ on Nov 20, 1979,
and $B = 17.6$ (a low state) on Apr 11, 1980. Since a typical $B-V$ value for MV Lyr is 
-0.25 (RRM),
we estimate $V = 13.6$ at the time of the {\it IUE} intermediate
state observations (Dec. 2, 1979), roughly a magnitude below typical high state values, and at the end
of the intermediate state. We calculate a $B$ magnitude change, low state-intermediate
state of ${\Delta}B$=3.7 magnitudes, with ${\Delta}V$ approximately the same. The 
\citet{hk04} MV Lyr light curve plot gives a
typical intermediate state--low state ${\Delta}V = 4.0$.

We have calculated an extension of the Fig.~5 model to 10,000\AA,
and have convolved the spectrum with the transmission profile of the $V$ filter \citep*{ms63}.
The normalized integral of this product tabulation is a measure of the theoretical $V$ magnitude.
This, together with a corresponding calculation with the low state spectrum, extended to 10,000\AA,
permits determination of the theoretical $V$ magnitude change between the low
and intermediate states. 
The Fig.~5 model produces ${\Delta}V = 4.91$, compared with the observed 3.7 magnitude change in $B$
(and $V$) between
Nov. 20, 1979 and Apr. 11, 1980 (RRM). 
We have calculated several models 
that fit the {\it IUE} spectra well, using elevated $T_{\rm eff}$ values in the outer
part of the accretion disk, and all of them produce a ${\Delta}V$ that is at least
one magnitude too large. 
If there is a bright spot in the intermediate state system the discrepancy is only made
worse by adding a model of it to our calculated system model.
This impasse indicates that a different approach is needed, and the black body plot in
Fig.~5 suggests that the difficulty is associated with the annulus synthetic spectra. 

\section{A limiting intermediate state model}

All of our models assume an outer boundary coincident with the tidal cutoff radius.
Terminating the outer accretion disk boundary at a smaller value has the prospective advantage
of reducing the flux at 5500\AA~while having a minor effect at 3000\AA. We have followed
this option to a limiting model: an isothermal 14,000K accretion disk terminated at
$r=20{\times}r_{\rm wd}$. This termination radius produces a flux level that matches the
observed spectrum, given the distance to MV Lyr.
We have preserved a truncation radius at $r=1.7{\times}r_{\rm wd}$,
although the flux contribution from the innermost annuli is small. The calculated
${\Delta}V$ for this model, low to intermediate state, is 4.06, in good agreement with
the \citet*{hk04} amplitude but larger than the RRM amplitude.
Fig.~6 shows the fit of 
this model to the {\it IUE} spectrum. The required flux level determines the accretion disk radius.
The radiation peak at 1300\AA~marks the same peak of the 14,000K source synthetic
spectrum of an annulus model. The fact that the limiting model synthetic spectrum is too blue is
a property of the Balmer continuum region of the source synthetic spectrum, beyond the
direct control of BINSYN. 
The black body plot of Fig.~5 suggests that, if formation of the
continuum takes place under conditions that reduce the Balmer jump and slope of the
Balmer continuum, then the Fig.~6 model might fit the {\it IUE} spectra closely. 
Reducing the $T_{\rm eff}$
of the isothermal accretion disk, to make it less blue, requires an increase in the
accretion disk radius, to provide the necessary flux. This increases the $V$ amplitude,
producing a disagreement with observation. Our limiting model represents the best fit to
the full array of observational data.

\begin{figure}
\figurenum{6}
\plotone{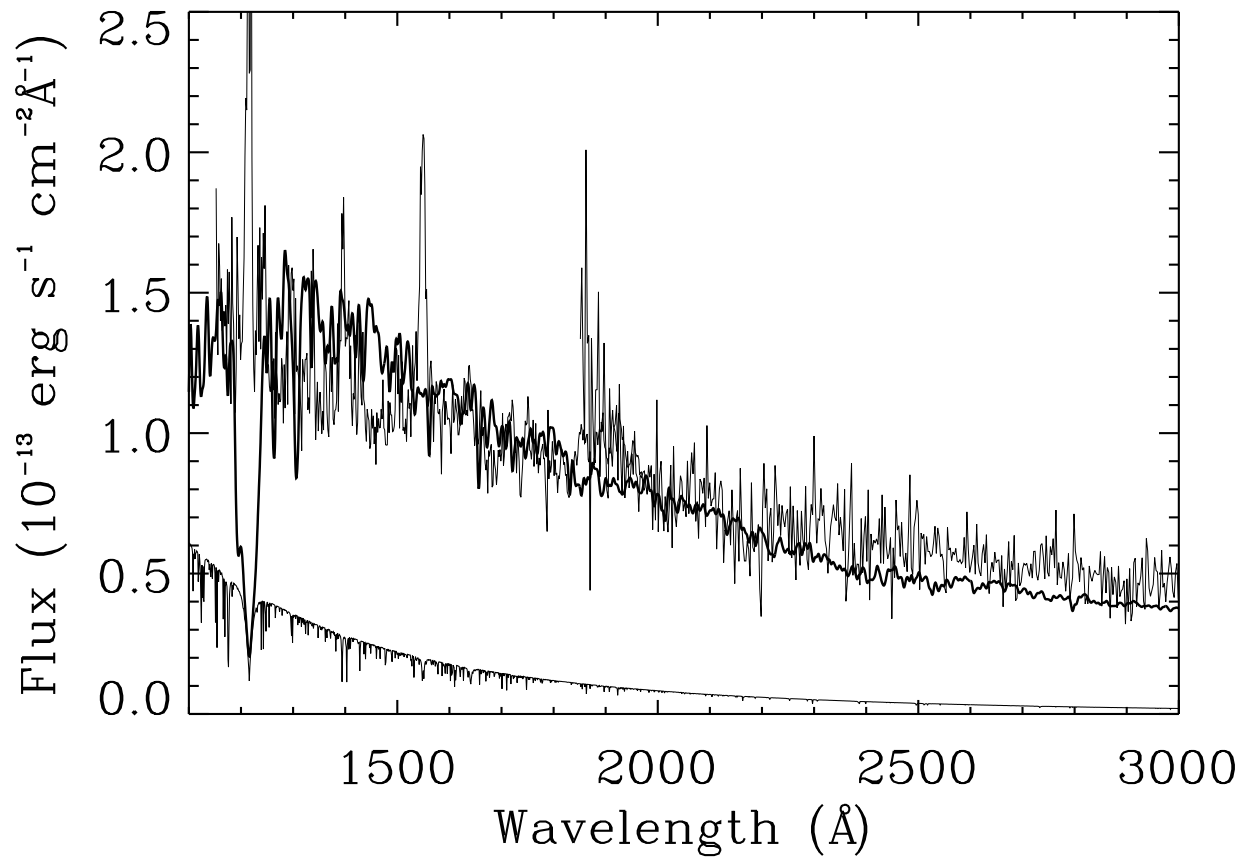}
\vspace{0.1cm}
\caption{As in Fig.~5 but with an isothermal 14,000K accretion disk 
extending from an inner truncation radius at $r=1.7{\times}r_{\rm wd}$
to $r=20.0{\times}r_{\rm wd}$. The light line is the {\it IUE} spectrum;
the heavy line (colored blue in electronic edition) is the synthetic spectrum.
The low state synthetic spectrum from Fig.~1 (lower spectrum) is shown for comparison.
{\it (See the electronic edition of the Journal for a color	version of this figure.)}
\label{fig6}}
\end{figure}

\section{Fit to the high state HST spectrum}

As part of an HST snapshot program, we obtained a 730 s exposure of MV Lyr on June 24, 2003
when it was in a peak high state of V$\sim$12.5. The Space Telescope Imaging Spectrograph
(STIS) was used with grating G140L, giving a resolution near 1.2\AA. Data reduction used
the latest release of CALSTIS (V2.13b). This version takes into account the decaying sensitivity
of the G140L grating.
All the lines are in
absorption (Fig.~7, discussed below), consistent with an optically thick disk. 

We investigated whether a Standard Model synthetic spectrum, including the option of a 
truncated accretion disk, can fit the HST spectrum. 
We
began with the L1999 mass transfer rate of 
$1.1{\times}10^{-8}{\cal M}_{\odot}{\rm yr}^{-1}$ and no truncation and found that
the predicted flux for this model is appreciably too large at all HST wavelengths.
The same mass transfer rate  
and an accretion disk truncated at 
$r/r_{\rm wd}=6.2$ leads to a synthetic spectrum that is too blue and a slightly too small
flux level, 
and an accretion disk truncated at $r/r_{\rm wd}=5.3$
produces a synthetic spectrum that is too blue and a slightly too large flux level.
Successively smaller mass transfer rates and no disk truncation produced models with flux
levels still too large but with a smaller mean residual. At each mass transfer rate it was possible
to find a pair of truncation radii for which the corresponding synthetic spectra bracketed
the HST spectrum, one with slightly too large flux values, the other too small; in each case,
both bracketing spectra were too blue. In addition to the initial test, we tested mass transfer rates of
$7.0{\times}10^{-9}{\cal M}_{\odot}{\rm yr}^{-1}$,
$5.0{\times}10^{-9}{\cal M}_{\odot}{\rm yr}^{-1}$,
$4.5{\times}10^{-9}{\cal M}_{\odot}{\rm yr}^{-1}$,
$4.0{\times}10^{-9}{\cal M}_{\odot}{\rm yr}^{-1}$, and finally
$3.0{\times}10^{-9}{\cal M}_{\odot}{\rm yr}^{-1}$.
In this latter case an
accretion disk extending to the WD surface fits the HST spectrum well between 1150\AA~and
1200\AA, but is much too blue (too little flux) longward of 1200\AA. Smaller mass transfer 
rates produce synthetic
spectra with too small flux at all HST wavelengths. We conclude that Standard Model synthetic 
spectra, with or without
truncated accretion disks, do not fit the HST spectrum; they are too blue.

With the objective to flatten the theoretical spectral gradient,
we selected the Standard Model with small errors at
the short wavelength extreme, mass transfer rate of $3.0{\times}10^{-9}{\cal M}_{\odot}{\rm yr}^{-1}$,
and modified the $T(R)$ relation to an isothermal outer region
with elevated temperature.	
We proceeded by initially selecting an 
accretion disk
annulus at a large radius and setting the $T_{\rm eff}$ values of all annuli of larger
radius to the $T_{\rm eff}$ of the selected annulus. The resulting synthetic spectrum provided
an improved fit to the HST spectrum. We continued, gradually moving to a smaller crossover
annulus and consequently higher $T_{\rm eff}$ value, until we reached a crossover annulus of 
$r/r_{\rm wd}=14.0$
and annuli $T_{\rm eff}=12,000$K at larger radii. At smaller radii the annulus $T_{\rm eff}$
values continued to follow a Standard Model for a mass transfer rate of
$3.0{\times}10^{-9}{\cal M}_{\odot}{\rm yr}^{-1}$.
The spectral gradient now was about correct but the overall flux level had become somewhat too large, 
and so we reduced
the flux level by truncating the accretion disk. Because of the high $T_{\rm eff}$ value
at the truncation radius, the color of the synthetic spectrum is mildly sensitive to the truncation radius value.
We found an optimum fit with a truncation
radius of $r/r_{\rm wd}=1.7$ and 
$T_{\rm eff}$ there of 43,404K.
It is likely that a slightly modified mass transfer rate, together with slightly changed truncation
radius and crossover radius and outer annulus $T_{\rm eff}$ value would fit the HST spectrum equally
well. For that reason, we do not attribute special significance to the fact that the truncation radius
found here is the same as the truncation radius for the intermediate state.
The large differences between the HST spectrum and the synthetic spectrum absorption lines lead to
larger mean and mean absolute residuals than for the intermediate and low states.
The mean residual of the fit to the 970 points between 1150\AA~and 1715\AA~was -0.44 and
the mean absolute residual was 0.71. These residuals apply to the scaled spectra.

\begin{figure}
\figurenum{7}
\plotone{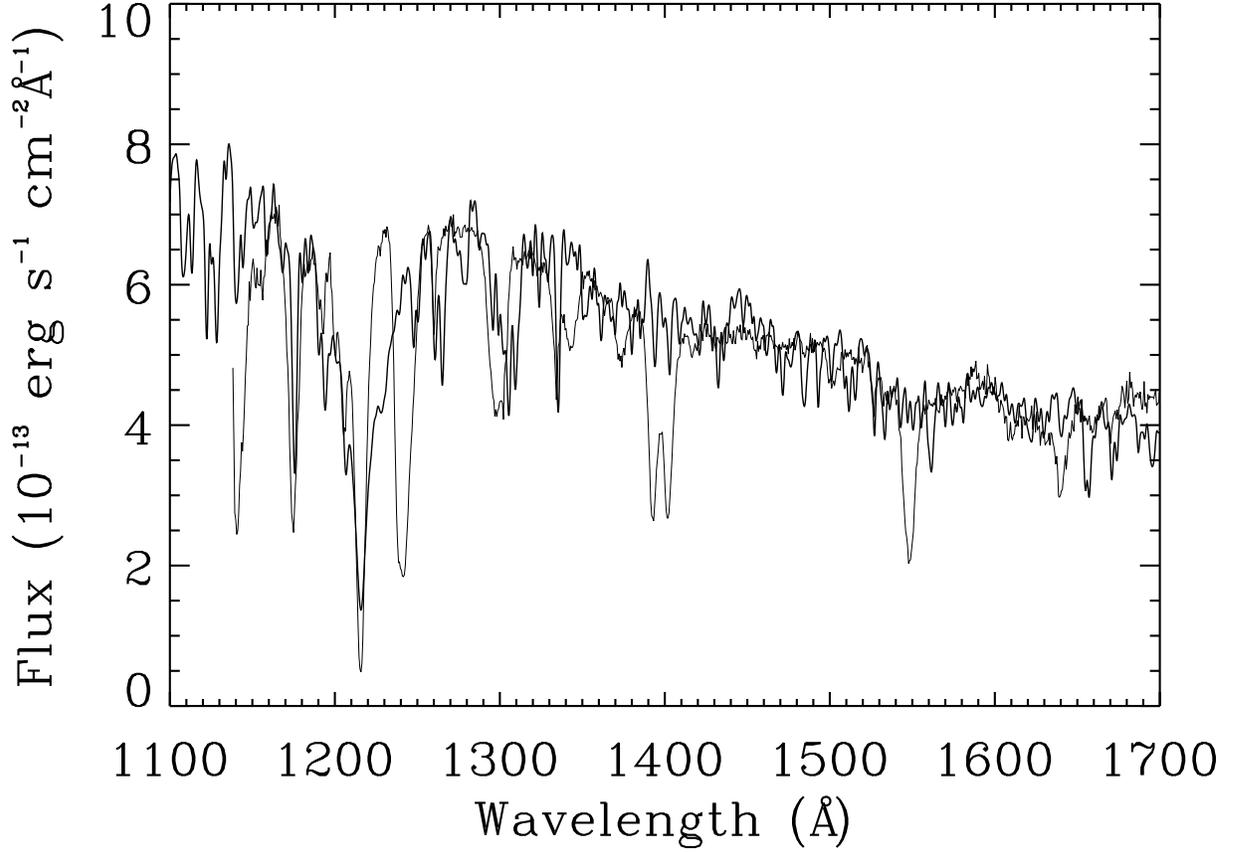}
\vspace{0.1cm}
\caption{Fit of a standard model
accretion disk with a mass transfer rate of
$3.0{\times}10^{-9}{\cal M}_{\odot}{\rm yr}^{-1}$, truncated at
an inner radius of $1.7{\times}r_{\rm wd}$, and with annuli $T_{\rm eff}$ values set to
12,000K beyond $r/r_{\rm wd}=14.0$ (thick line, colored blue in electronic edition) to an HST 
high state spectrum of MV Lyr
(light line).
The synthetic spectrum has been convolved with a Gaussian 1.2\AA~FWHM broadening
function. Line identifications are in following plots.
{\it (See the electronic edition of the Journal for a color	version of this figure.)}
\label{fig7}}
\end{figure}

\begin{figure}
\figurenum{8}
\plotone{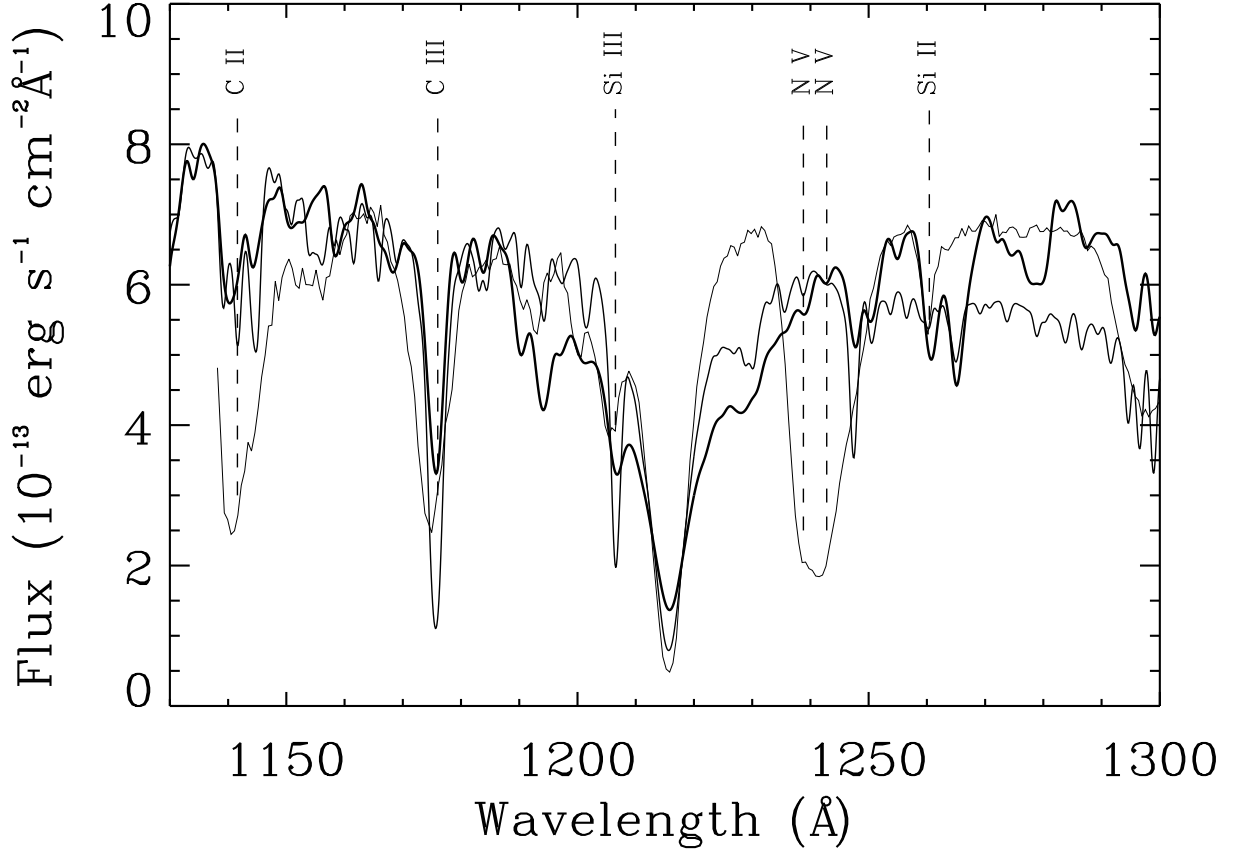}
\vspace{0.1cm}
\caption{Detail of Fig.~7. The intermediate strength overplotted spectrum (colored red 
in electronic edition)
is a synthetic spectrum 
of a single annulus with a $T_{\rm eff}$ of 31,564K, convolved with a Gaussian 1.2\AA~FWHM 
broadening function. 
Note the much closer fit to
the observed Ly ${\alpha}$ (light line) than provided by the system synthetic spectrum (heavy 
line, colored blue in electronic edition). 
The annulus synthetic spectrum was divided by $6.5{\times}10^{8}$ before overplotting, to place
the synthetic spectrum over the observed spectrum. 
{\it (See the electronic edition of the Journal for a color	version of this figure.)}
\label{fig8}}
\end{figure}

\begin{figure}
\figurenum{9}
\plotone{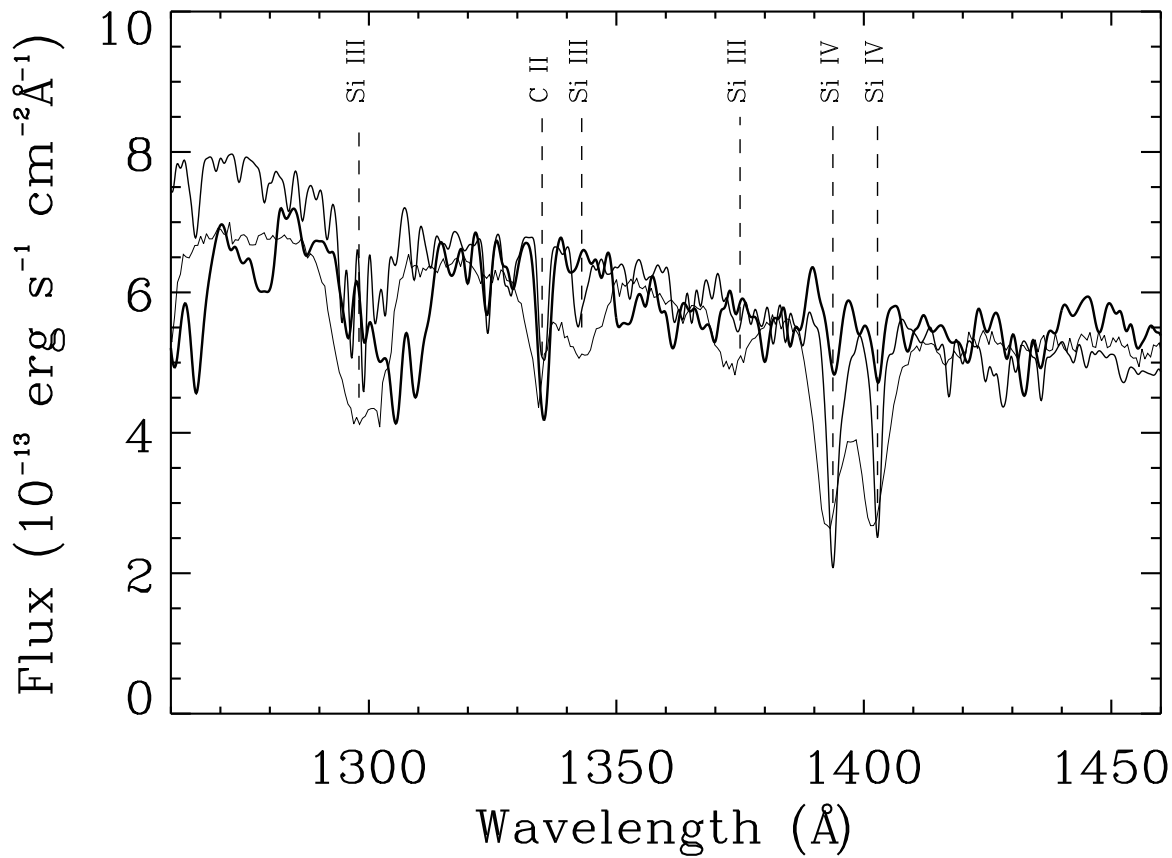}
\vspace{0.1cm}
\caption{As in Fig.~8, for a different spectral region.	The C II and Si III labels
mark groups of transitions. Note that the Si IV equivalent widths in the overplotted annulus
spectrum (31,564K; intermediate strength line, colored red in electronic edition) are much greater than 
for the system synthetic 
spectrum (heavy line, colored blue in electronic edition), indicating 
that the HST 
spectral lines (light plotted line) are formed in a higher temperature 
environment than for the system
synthetic spectrum.	Note also that the observed Si IV lines are broadened as well as
displaced to shorter wavelengths.
The annulus synthetic spectrum was divided by $4.7{\times}10^{8}$ before overplotting, to place
the synthetic spectrum over the observed spectrum.
{\it (See the electronic edition of the Journal for a color	version of this figure.)}
\label{fig9}}
\end{figure}

\begin{figure}
\figurenum{10}
\plotone{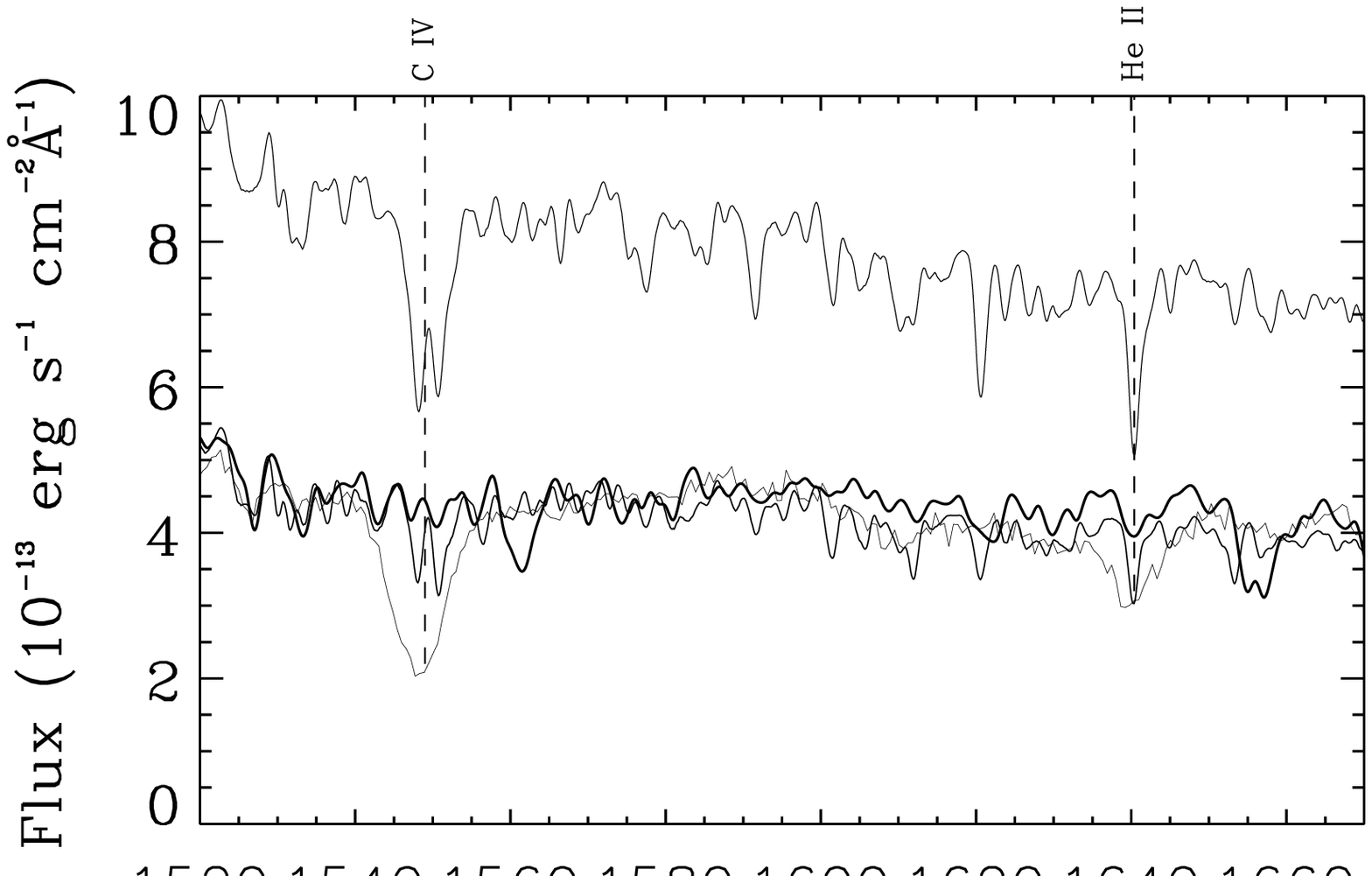}
\vspace{0.1cm}
\caption{As in Fig.~8. Note that the equivalent widths of the C IV doublet and the He II
line are appreciably greater in the overplotted annulus spectrum (31,564K; intermediate strength line,
colored red in electronic edition)
than in the system synthetic 
spectrum (heavy line, colored blue in electronic edition). The spectrum at the top, colored green
in the electronic edition, represents a
38,076K annulus, shown for comparison, and with the same normalizing factor as the 31,564K annulus. 
The annulus spectra were divided by $3.9{\times}10^{8}$
before overplotting, to place the 31,546K synthetic spectrum over the observed spectrum.
When scaled to fit the
observed spectrum, the 38,076K spectrum is fairly similar to the 31,546K spectrum.
Note that the C IV doublet is resolved in both annulus spectra
and in the system synthetic spectrum. All three of these spectra have been convolved with a
1.2\AA~Gaussian broadening function. 
{\it (See the electronic edition of the Journal for a color	version of this figure.)}
\label{fig10}}
\end{figure}

Figures 7--10 show the fit to the HST spectrum
of a synthetic spectrum with our optimum high state parameters.
Although the continuum is a reasonable fit,
individual line fits show appreciable discrepancies, especially
for the high excitation lines of NV, Si IV, C IV, and He II.
Since all of our synthetic spectra assume solar composition,
the discrepancies could arise from: (1) abundance differences from
solar composition, or (2) excitation differences
between the formation regions for the absorption lines and the continuum. 

The observed absorption lines show a slight blue shift corresponding to a velocity
of about $-200\,\mathrm{km\,s^{-1}}$.
This effect is apparent in the detail plots, Fig.~8 (C III), Fig.~9 (C II, Si IV), 
and Fig.~10 (C IV, He II).  
\citet*{sch81} and SPT95 separately determine a system radial velocity of MV Lyr
of $\gamma~{\simeq}~-35\,\mathrm{km\,s^{-1}}$ from optical wavelength spectra.
The velocity difference (of $165\mathrm{km\,s^{-1}}$) is a clear indication that 
the absorption lines are formed
in a wind.
Since MV Lyr is viewed nearly pole-on, this means the wind
has a low velocity. This interpretation is consistent with the observed line broadenings
as listed in Table~2,
e.g., the Si IV doublet near 1400\AA.

Fig.~8 shows the vicinity of Ly ${\alpha}$. 
Note the much better
fit of the overplotted annulus synthetic spectrum ($T_{\rm eff}=31,564$K)
to the observed Ly ${\alpha}$ line 
than is true
for the system synthetic spectrum, whose mean $T_{\rm eff}$ is appreciably lower.
This indicates that the absorption lines are formed in a higher temperature environment
(i.e., corona or wind)
than is true for the continuum. 
Fig.~9 shows the Si IV doublet near 1400\AA~; it
is too weak in the system synthetic spectrum, and equivalent widths for the
overplotted annulus synthetic spectrum ($T_{\rm eff}=31,564$K) are again a better fit. 

The C IV doublet (Fig.~10) in the
system synthetic spectrum is also much too weak. The doublet is stronger in both the overplotted 
annulus synthetic
spectra ($T_{\rm eff}$=31,564K, and 38,076K), but not by enough when scaled to fit the observed 
profile. The 
C II line complex 
(Fig.~9) is slightly too weak
in the overplotted annulus spectrum, while the equivalent width of the system
synthetic spectrum appears comparable to the observed value.
The C III line complex, Fig.~8 in the
overplotted annulus synthetic spectrum, appears to have an equivalent width approximately 
matching
the observed value, while the system synthetic spectrum is too weak. 
The varying strengths of C lines in different
ionization stages
and the observed (large) strength 
of N V (Fig.~8) may be due to composition effects but
we cannot exclude
the possibility that it is an excitation effect.
Large N V/C IV ratios have been seen in several CV systems
\citep{G2003} and are likely related to evolutionary effects. Note that H2004 determines
an overall metal composition for the MV Lyr WD of $Z=0.3{\times}Z_{\odot}$, but detailed
elemental abundances, relative to solar values, of C=0.5, N=0.5, and Si=0.2. (Since strong
excitation effects are clearly present there is no conflict between our assumed solar
abundance for the present analysis and the WD composition determination in H2004.)

It is of interest to compare the emission line spectrum of the intermediate state
and the absorption line spectrum of the high state (Fig.~11).
Note that emission lines in the intermediate state match
corresponding absorption lines in the high state. 
As well as can be judged,
the
relative strengths of the various resonance emission lines are similar to the relative strengths
of the same absorption lines. The HST spectrum in Fig.~11 has been displaced 
downward to 
superpose
the C IV complex. The two broadened spectra fit accurately at the superposed line boundaries; 
if there were a radial velocity
difference between the emission line spectrum and the absorption line spectrum as large as
the displacement of the emission line spectrum (e.g., Fig.~9) from the synthetic spectrum,
that difference would be detectable. Comparable fits result when other spectral regions
are superposed, although the low signal level of the {\it IUE} spectrum produces some
ambiguity. 
It is of particular interest that the intermediate state spectrum is flatter than the 
high state spectrum.
The two sets of lines in the intermediate and high states appear to be produced
in a wind with a similar velocity structure in the two states.

\begin{figure}
\figurenum{11}
\plotone{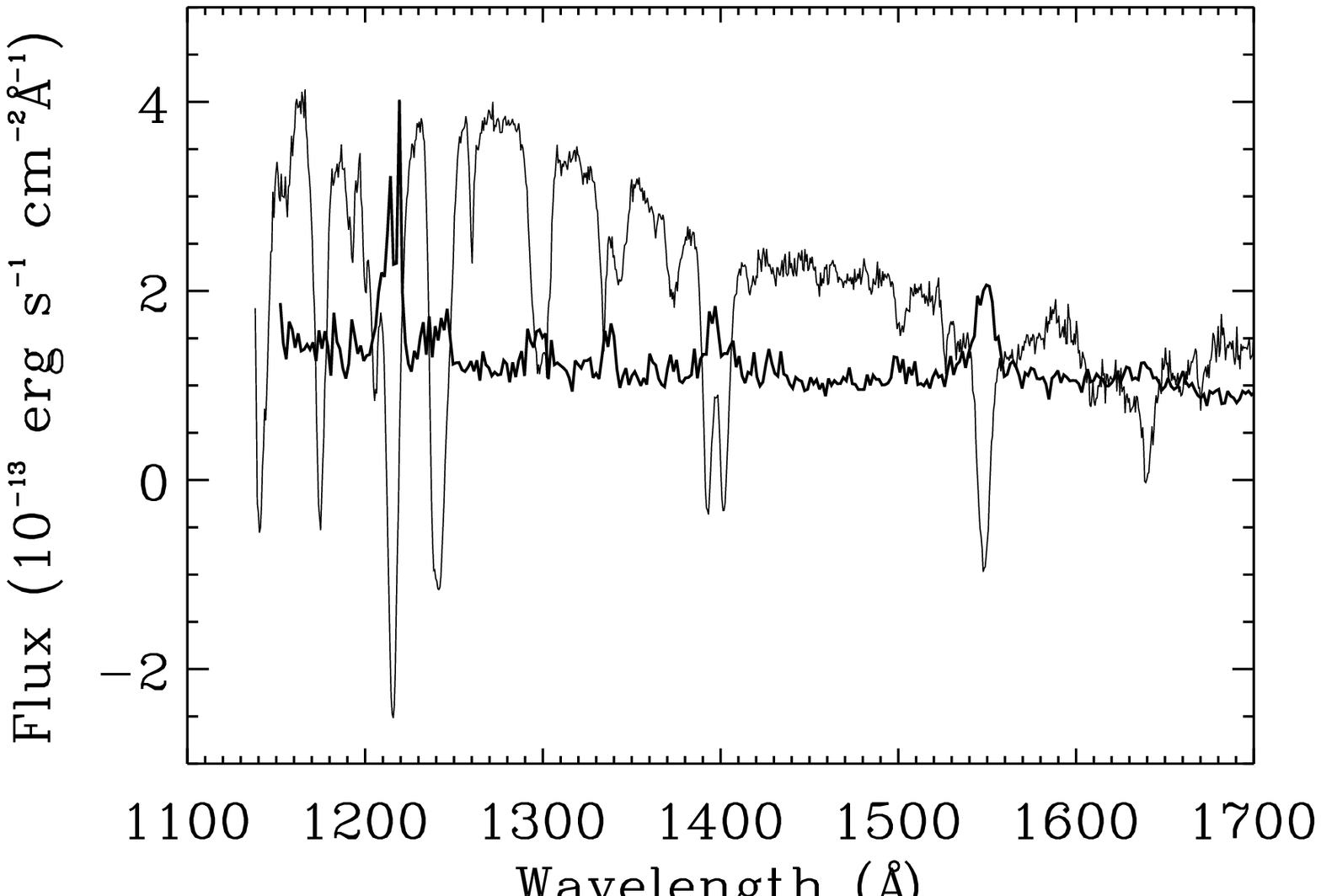}
\vspace{0.1cm}
\caption{Comparison of the high state HST spectrum (upper spectrum) and the intermediate
state {\it IUE} spectrum (lower spectrum) of MV Lyr. The high state spectrum has been displaced 
downward by 
3.0 flux units (ordinate scale) to superpose the C IV complex at 1540\AA.
This plot demonstrates that the strong emission lines in the {\it IUE} spectrum become strong
absorption lines in the HST spectrum, at (as well as can be determined) the same wavelengths
and with similar broadening.
A wavelength displacement as large as that between the HST spectrum and the synthetic spectrum 
in
Fig.~9 would be detectable. Both sets of lines appear to be produced in a low velocity wind.} 
 
\label{fig11}
\end{figure}

\begin{deluxetable}{lclllcl}
\tablewidth{0pt}
\tablecaption{Prominent absorption and emission lines, FWHM values}
\tablehead{
\colhead{ion} & \colhead{nominal ${\lambda}$(\AA)} & \colhead{HST} 
& \colhead{{\it IUE}}  & \colhead{low}  & \colhead{intermed.} & \colhead{high}}
\startdata
C II       &  1141	     & 10.5      &   \nodata    &  0.5  &  5.8  &  4.4\\
C III      &  1175	     & 6.3       &   1.4    &  2.6  &  2.6  &  3.2\\
N V        &  1241	     & 11.5      &   1.2    &  2.6  &  1.0  &  1.8\\
Si II      &  1260       & 2.4       &   1.2    &  0.3  &  8.0  &  3.9\\
Si III     &  1299       & 12.3      &   2.0    &  2.2  &  1.7  &  4.8\\
C II       &  1334       & 3.0       &   7.5\tablenotemark{a} \tablenotemark{b}    &  1.0  &  2.8  &  2.7\\
Si III     &  1342		 & 9.9       &          &  1.0  &  2.7  &  3.1\\
Si III     &  1374       & 8.6       &   1.5    &  1.6  &  1.3  &  1.6\\
Si IV      &  1393       & 6.4       &   7.0\tablenotemark{a} \tablenotemark{b}    &  1.8  &  2.0  &  3.0\\
Si IV      &  1402       & 7.4       &          &  1.7  &  2.0  &  3.0\\
Si III     &  1501       & 6.6       &   7.0\tablenotemark{a}    &  0.8  &  2.1  &  2.4\\
C IV       &  1549       & 7.9       &   10.8\tablenotemark{a}   &  4.5  &  3.2  &  5.1\\
He II      &  1640       & 6.6       &   7.0\tablenotemark{a}    &  3.4  &  3.0  &  3.4\\
\enddata
\tablenotetext{a}{emission line}
\tablenotetext{b}{this entry is the FWHM sum with the ion on the following line}
\tablecomments{The {\it IUE} column refers to the intermediate state spectrum, SWP07296. 
The columns
designated as "low", "intermed.", and "high" refer to the system synthetic spectra corresponding
to Fig.~1, Fig.~5, and Fig.~7 respectively.}
\end{deluxetable}

\section{Change in system luminosity, low state to high state}

The calculated
$V$ light change for our optimum model (Fig.~7), from low to high state, 
following the procedure of \S 5, is 5.91 magnitudes. 
The observed values are; AAVSO, 6.0, and \citet*{hk04}, 5.25, bracketing our calculated
value.
The HST observations provide a separate test.
We used the STIS/CCD acquisition image of MV Lyr to determine the magnitude at the time
of the HST FUV observations to be 12.4 in the F28x50LP filter, corresponding to a
monochromatic flux (but integrated over an extended spectral region) of 
$2.205{\pm}0.05{\times}10^{-14}\,\mathrm{erg\,cm^{-2}s^{-1}\AA^{-1}}$
at 7228.5\,\AA. The filter has a lower transmission limit at 5400\AA, a peak response at 
6000\AA, and a gradual dropoff with half-peak response at 8300\AA~and an upper
transmission cutoff at 10,000\AA. 
The high state model synthetic spectrum, when multiplied by this filter transmission 
and integrated, produces a calculated equivalent monochromatic
7228.5\AA~flux, at the Earth, for the 505pc (H2004) distance to MV Lyr, of
$2.91{\times}10^{-14}\,\mathrm{erg\,cm^{-2}s^{-1}\AA^{-1}}$, in approximate agreement 
with the observed value.

\section{Discussion}

The significant result from our low state study is that, to within observational error,
the system spectrum can be represented by a combination of a WD spectrum and an irradiated, 
$T_{\rm eff,pole}=2600$K
secondary component. 
There is no observational evidence for accretion disk emission in the
MV Lyr low state. 
Our newly calculated contribution
of the secondary star is an interpolation among NextGen models and includes allowance for
irradiation of the secondary by the WD. 
Both L1999 and HL02 model the low state of MV Lyr with a truncated
accretion disk that retains most of its mass from the high state, and emits radiation
at a level consistent with residual viscosity. 
Our results limit possible low state accretion disk emission to values well below those 
predicted by both L1999 and HL02. \citet*{hcr94} discuss the occasional sudden drops in
luminosity of V974 Aql and show that the drops can be explained within the accretion
disk limit cycle theory in terms of back and forth propagation of cooling and heating
transition fronts. As with L1999 and HL02, \citet*{hcr94} find that the accretion disk
undergoes little net mass loss while mass transfer from the secondary is turned off.

HL02 discuss slow passage of an accretion disk system from an initial stable state
through the
instability region and show that outside-in outbursts would occur but are
unobserved. 
The only solution to the VY Scl phenomenon that they visualize,
to meet the constraint imposed by the case of TT Ari \citep{g1999}, is a
magnetic field sufficiently large (B${\gtrsim}6$MG) to truncate the 
residual low state
accretion disk
at a radius large enough to suppress accretion disk formation. However, as HL02 note, 
if a magnetic field is
present, then circular polarization should also be present. \citet{r1981} chronicles that
repeated attempts to measure the 
circular polarization 
in MV Lyr always
found a value less than 0.13\%.
Robinson et al. also note that, if there is a magnetic field present that is strong
enough to disrupt the accretion disk, Zeeman-splitting of absorption lines should
be detected; it is not.	Thus, there is no evidence for a strong magnetic field associated with
the MV Lyr WD.

Both the intermediate state {\it IUE} spectra and the high state HST spectra are
inconsistent with our calculated Standard Models, having flatter flux profiles.
Our subsequent thin disk simulations, using TLUSTY annulus models, differ from the
Standard Model by assuming a flatter $T(R)$ profile;
they are able to produce
fairly good fits to the observed continua.
Confirmation that the TLUSTY annuli represent a thin disk 
model is provided by the output data for an annulus at $r/r_{\rm wd}=18.5$ and for a
mass transfer rate of $4.0{\times}10^{-9}{\cal M}_{\odot}{\rm yr}^{-1}$, close to the value
of our high state model.
For this annulus, whose radius is $1.45{\times}10^{10}$ cm, 
a Rosseland optical depth of 0.81 is located at
$z=1.94{\times}10^9$ cm, one dex
smaller than the radius. A Rosseland
optical depth of 0.001 occurs at $z=1.98{\times}10^9$ cm, indicating line formation in
a thin photosphere.

Smak (1994) has discussed peculiar (i.e., non-Standard Model),
$T(R)$ distributions.
The departures from the Standard Model in the systems he cites (e.g., \citealt*{rpt92}) 
are in the direction 
of a too flat 
$T(R)$ profile, as we find for MV Lyr. 
Smak argues that the departures for systems with $i>75{\arcdeg}$
result from the assumption of a flat accretion disk. That explanation is not applicable
for MV Lyr
because of its low inclination. 
Smak suggests that, for low $i$ cases, heating of the outer part of the accretion 
disk by the stream collision could be the explanation. Buat-M\'{e}nard, Hameury \& Lasota (2001)
discuss the effect of stream impact heating and tidal effects. 
As our high state model indicates, if stream impact heating 
is the physical
cause of anomalous heating in the outer accretion disk region, the heating effects extend over
an appreciable fraction of the accretion disk radius. This effect differs from a bright spot,
for which the radial extent amounts to a few percent of the component separation 
\citep[see][sect.2.6.5]{w95}.
We stress the importance of the low orbital inclination in MV Lyr and the consequential absence 
of eclipses. 
All of the systems cited
by Smak were shown to have peculiar $T(R)$ distributions by application of the MEM technique
(\citealt{h93}) and so were restricted to systems showing eclipses. The spectrum synthesis
method used in this paper represents an alternative and independent technique that is
applicable to both eclipsing and non-eclipsing systems. 
 
The presence of high-excitation absorption lines in the MV Lyr system
imply their formation in a high temperature region above (both faces of) the accretion disk, possibly
similar to a chromosphere or corona.	
Accretion disk coronae are considered in \citet*{MMH89,MMH94}, \citet{L97}, and \citet{h2000}.
We have presented evidence
for formation of both the intermediate state emission lines and the high state absorption lines 
in a wind, apparently coextensive with the high temperature region.
It would be of interest to
search for a P-Cygni profile, possibly in the infra-red. We thank the referee for this suggestion.
\citet{d1997} summarizes known properties of winds from CVs. \citet*{psd98} discuss a
model of radiation-driven winds from accretion disks, and
\citet*{p04} discuss the stability of line-driven winds. 
Table~2 lists the FWHM wavelength ranges for the principal absorption lines in the HST spectrum.
In most cases, several transitions contribute to a given line. 
The low MV Lyr $i$ value nominally would imply
narrow disk lines, distinctly different from the observations. The observed broadening 
is consistent with lines of sight that traverse
a range of radial velocities before reaching an optical depth of 1.0.
The $200\,\mathrm{km\,s^{-1}}$ blue shift of
the absorption lines is smaller than typically found for winds from accretion disks.

Our TLUSTY annuli models are in LTE and apply default values for the internal parameters,
leading to a negative temperature gradient perpendicular to the surface of the annulus.
Line formation in a vertically extended wind violates our model assumption of hydrostatic
equilibrium; addition of a wind model, 
\citep*{lk02}, might provide improved representation of the MV Lyr spectral data. \citet*{f2002}
combined TLUSTY disk models with a representation of a biconical wind in a study of SS Cyg.

We speculate that continuum formation takes place under conditions different from the
assumptions of our annulus models and that difference may explain our failure to
achieve an accurate continuum fit, especially in the intermediate state.
\citet{h1989a,h1989b} discusses	conditions for hot upper layers of accretion disk annuli to exist. 
He shows 
that the z-direction profile of viscosity strongly affects the formation of these regions.
\citet*{hh98} discuss the treatment of viscosity as implemented in TLUSTY, in particular
the steps taken to prevent a high temperature z-direction anomaly. The authors also emphasize
the importance of NLTE effects.
It is widely believed that the physical cause of accretion disk
viscosity is the magnetorotational instability \citep*{bh91}. \citet{b02} discusses issues
in the implementation of the instability in accretion disk models. The magnetohydrodynamical
simulations of \citet*{s96} find an increase in viscosity toward the accretion disk surface,
leading to increased z-direction heating. 

The flatter spectral profile of the intermediate state {\it IUE} spectrum, as compared to
the HST high state spectrum (Fig.~11) is of particular interest. 
The isothermal accretion disk extending only half way to the tidal cutoff radius in the
intermediate state has 
the attractive feature that it provides a ready
explanation for emission lines (the line emission region is
seen extending beyond the edge of the accretion disk), while the same lines are seen in
absorption in the high state (the accretion disk now extends to the tidal cutoff radius).
This scenario indirectly supports the absence of an accretion disk
in the low state, again in disagreement with L1999 and HL02. 
An explanation of the flat $T(R)$ profile, both intermediate and high, must differ from the case of
DW UMa \citep{kn00,a-b03}, also a VY Scl star. In DW UMa there is also a flat $T(R)$ profile,
but it is explained by self-occultation by a puffed-up outer accretion disk rim and an orbital
inclination large enough to produce eclipses of the WD.

\section{Summary}

(1) Based on
the excellent fit of a system synthetic spectrum to FUSE, {\it IUE}, and optical 
low state spectra (H2004),
any accretion disk that may be present in the low state must have a $T_{\rm eff}$
less than 2500K.
This result agrees with a study of the CV system TT Ari \citep*{g1999}. A comparison of our results
(Fig.~3) with two models (L1999 and HL02) of the VY Scl phenomenon demonstrates
a discrepancy with those models.

(2) Using TLUSTY annulus models, and corresponding synthetic spectra, to simulate an
intermediate state for which {\it IUE} spectra are available, we show that a Standard
Model (see text) cannot produce a system synthetic spectrum that satisfactorily fits the 
{\it IUE}
spectra. 
A rough fit is possible with a Standard Model with a mass transfer rate of
$1.0{\times}10^{-9}{\cal M}_{\odot}{\rm yr}^{-1}$ but the model synthetic spectrum is
too blue.
A fairly good fit is possible with a mass transfer rate of
$6.{\times}10^{-10}{\cal M}_{\odot}{\rm yr}^{-1}$,
a truncation radius of $r=1.7{\times}r_{\rm wd}$, and with $T(R)$=9500K beyond 
$r=11.5{\times}r_{\rm wd}$.	But this model produces a change in $V$ magnitude between
the low and intermediate states that is about one magnitude larger than the
observational value. An isothermal 14,000K accretion disk with a radius that is 
half of the tidal cutoff radius produces agreement with the V magnitude change; 
this model is slightly bluer than the {\it IUE} spectra.

(3) Study of a high state demonstrates that no
combination of mass transfer rate and truncation radius can produce a 
Standard Model system
synthetic spectrum whose continuum satisfactorily fits our recent HST spectrum.
All synthetic spectra are too blue.
A fairly good fit is possible
for a mass transfer rate of $3.0{\times}10^{-9}{\cal M}_{\odot}{\rm yr}^{-1}$, a
truncation radius of $r=1.7{\times}r_{\rm wd}$, and $T(R)$=12,000K beyond 
$r=14.0{\times}r_{\rm wd}$. This model produces a calculated system flux at 7228.5\AA~that
is in reasonable agreement with a measurement of HST optical flux.

(4) The absorption lines in our MV Lyr HST spectrum show a range of excitations, and corresponding
temperatures, that are higher than the nearly flat continuum would imply. This result is 
consistent with line formation in a large $z$ high temperature region. Anomalous line strengths of 
C, N, Si, and He
have a possible interpretation in terms of composition effects, but excitation effects
cannot be disentangled.

(5) The slight blue shift of the absorption lines (${\sim}-200\,\mathrm{km\,s^{-1}}$), 
as compared with the system
radial velocity of $-35\,\mathrm{km\,s^{-1}}$, indicates that the absorption lines 
are formed in a
low velocity wind, which is coextensive with the large $z$ high temperature region.
There is no apparent difference in velocity between the emission line intermediate state
spectrum and the high state absorption line spectrum; the wind and high temperature structures
appear to be present in both the intermediate and high states. 

(6) The difference between the physical conditions under which the absorption 
and emission lines form
in a wind
and the physical conditions assumed by our hydrostatic equilibrium annulus models, 
used to calculate the synthetic 
spectra, may explain
our difficulty in fitting the high state and intermediate state continua accurately.
The same problem may explain the failure of our synthetic spectra to represent the
high excitation absorption lines in the high state HST spectra.

\section{Acknowledgments}
We thank the anonymous referee for detailed comments which improved the presentation
of this paper.
APL and PS are grateful for partial support from NASA Grant GO-9357 from the
Space Telescope Science Institute and for FUSE grant NAG 5-12203. 
PS, KSL and EMS also acknowledge support for this work provided by NASA through
grants GO-9357 and GO-9724 from the Space Telescope Science Institute, which is
operated by AURA, Inc., under NASA contract NAS 5-26555.
BTG was supported 
by a PPARC Advanced Fellowship.

\clearpage


\clearpage

\begin{thebibliography}{}
\bibitem[Andronov, Pinsonneault, \& Sills (2003)]{aps03}
Andronov, N., Pinsonneault, M., \& Sills, A. 2003, \apj, 582, 358
\bibitem[Araujo-Betancor et al.(2003)]{a-b03}
Araujo-Beatncor, S., Knigge, C., Long, K.S., Hoard, D.W., Szkody, P., Rodgers, B., Krisciunas, K., Dhillon, V.S.,
Hynes, R.I., Patterson, J., \& Kemp, J. 2003, \apj, 583, 437
\bibitem[Armitage \& Livio (1998)]{al98}
Armitage, P.J., \& Livio, M. 1998, \apj, 493, 898
\bibitem[Balbus (2002)]{b02}
Balbus, S.A. 2002, in The Physics of Cataclysmic Variables and Related Objects, ASP Conference
Series, Vol. 261, eds. B.T. G\"{a}nsicke, K. Beuermann, \& K. Reinsch,
(San Francisco:Astronomical Society of the Pacific), p.356
\bibitem[Balbus \& Hawley (1991)]{bh91}
Balbus, S.A., \& Hawley, J.F. 1991, \apj, 376, 214
\bibitem[Buat-M\'{e}nard, Hameury \& Lasota(2001)]{bm2001}
Buat-M\'{e}nard, V., Hameury, J.-M., \& Lasota, J.-P. 2001, \aap, 366, 612
\bibitem[Drew (1997)]{d1997}
Drew, J.E. 1997, in Accretion Phenomena and Related Outflows, IAU Colloquium 163,
ASP Conference Series, Vol. 121 eds. D.T. Wickramasinghe, L. Ferrario, \& G.V. Bicknell
(San Francisco:Astronomical Society of the Pacific), p.465
\bibitem[Frank, King, \& Raine (1995)]{fkr95}
Frank, J., King, A., \& Raine, D. 1995, Accretion Power in Astrophysics 
(Cambridge:Univ. Press) 2nd ed., p.78 (FKR)
\bibitem[Froning et al.(2002)]{f2002}
Froning,C.S., Long, K.S., Drew, J.E., Knigge, C., Proga, D., \& Mattei, J.A.
2002, in The Physics of Cataclysmic Variables and Related Objects, ASP Conference Series, Vol. 261
eds. B.T. G\"{a}nsicke, K. Beuermann, \& K. Reinsch, (San Francisco:Astronomical Society of the Pacific), p.337
\bibitem[G\"{a}nsicke et al. (1999)]{g1999}
G\"{a}nsicke, B.T., Sion, E.M., Beuermann, K., Fabian, D., Cheng, F.H., and
Krautter, J. 1999, \aap, 347, 178
\bibitem[G\"{a}nsicke et al. (2003)]{G2003}
G\"{a}nsicke, B.T.,Szkody, P., de Martino, D., Beuermann, K., Long, K.S., Sion, E.M.,
Knigge, C., Marsh,T., \& Hubeny, I. 2003, \apj, 594, 443
\bibitem[Hamada \& Salpeter (1961)]{hs61}
Hamada, T., \& Salpeter, E.E. 1961, \apj, 134, 683
\bibitem[Hameury et al. (1988)]{h1988}
Hameury, J.-M., King, A.R., Lasota, J.-P., \& Ritter, H. 1988, \mnras, 231, 535
\bibitem[Hameury et al. (2000)]{h2000}
Hameury, J.-M.,Lasota, J.-P., \& Warner, B. 2000, \aap, 353, 244
\bibitem[(2002)]{hl02}
Hameury, J.-M. \& Lasota, J.-P. 2002, \aap, 394, 231(HL02)
\bibitem[Hauschildt et al. (1999)]{haus99}
Hauschildt, P.H., Allard, F., \& Baron, E. 1999, \apj, 512, 377
\bibitem[Hoard et al. (2004)]{h2004}
Hoard, D.W., Linnell, A.P., Szkody, P., Fried, R.E., Sion, E.M., Hubeny, I., \&
Wolfe, M.A. 2004, \apj, 604, 346 (H2004)
\bibitem[Honeycutt, Cannizzo, \& Robertson (1994)]{hcr94}
Honeycutt, R.K., Cannizzo, J.K., \& Robertson, J.W. 1994, \apj, 425, 835
\bibitem[Honeycutt \& Kafka (2004)]{hk04}
Honeycutt, R.K., \& Kafka, S. 2004, \aj, 128, 1279
\bibitem[Horne (1993)]{h93}
Horne, K. 1993, in Accretion Disks in Compact Stellar Systems, ed. J.C. Wheeler
(Singapore:World Scientific), p.117
\bibitem[Hubeny (1988)]{h88}
Hubeny, I. 1988, Comp. Phys. Comm., 52, 103
\bibitem[Hubeny (1989a)]{h1989a}
Hubeny, I. 1989a, in Theory of Accretion Disks, eds. F. Meyer, W. Duschl, J. Frank,
\& E. Meyer-Hofmeister (Dordrecht:Kluwer), p.445
\bibitem[Hubeny (1989b)]{h1989b}
Hubeny,I. 1989b, in Algols, ed. A.H. Batten (Dordrecht:Kluwer), p.117
\bibitem[Hubeny \& Lanz (1995)]{hl95}
Hubeny, I., \& Lanz, T. 1995, \apj, 439, 875
\bibitem[Hubeny, Stefl \& Harmanec (1985)]{hsh85}
Hubeny, I., Stefl, S., \& Harmanec, P. 1985, Bull. Astron. Inst. Czechosl. 36, 214
\bibitem[Hubeny \& Hubeny (1998)]{hh98}
Hubeny, I., \& Hubeny, V. 1998, \apj, 505, 558
\bibitem[King (1988)]{k1988}
King, A.R. 1988, Q.Jl R. astr. Soc., 29, 1
\bibitem[King \& Cannizzo (1998)]{kc98}
King, A.R., \& Cannizzo, J.K. 1998, \apj, 499, 348
\bibitem[King \& Schenker (2002)]{ks02}
King, A.R., \& Schenker, K. 2002, in The Physics of Cataclysmic Variables and Related Objects,
ASP Conference Series, Vol. 261, eds. B.T. G\"{a}nsicke, K. Beuermann, \& K. Reinsch 
(San Francisco:Astronomical Society of the Pacific), p.233
\bibitem[Knigge et al.(2000)]{kn00}
Knigge, C.,Long, K.S., Hoard, D.W., Szkody, Paula, \& Dhillon, V.S 2000, \apjl, 539, L49
\bibitem[K\v{r}i\v{z} \& Hubeny (1986)]{kh86}
K\v{r}i\v{z}, S. \& Hubeny, I. 1986, Bull. Astron. Inst. Czech., 37, 129
\bibitem[Lasota(2001)]{l2001}
Lasota, J.-P. 2001, New Astr. Rev., 45, 449
\bibitem[(1999)]{l1999}
Leach, R., Hessman, F.V., King, A.R., Stehle, R., \& Mattei, J.
1999, \mnras, 305, 225 (L1999)
\bibitem[Linnell \& Hubeny (1996)]{lin96}
Linnell, A.P., \& Hubeny, I. 1996, \apj, 471, 958
\bibitem[Linnell et al. (2005)]{l2005}
Linnell, A.P., Hoard, D.W., Szkody, Paula, \& Fried, R.E., 2005, in Astrophysics in the Far Ultraviolet:
Five Years of Discovery with FUSE, ASP Conf. Series, eds. G. Sonneborn, W. Moos, \& B-G Andersson
(San Francisco:Astronomical Society of the Pacific), submitted
\bibitem[Liu, Meyer \& Meyer-Hofmeister (1997)]{L97}
Liu, B.F., Meyer, F., \& Meyer-Hofmeister, E. 1997, \aap, 328, 247
\bibitem[Livio \& Pringle (1994)]{lp94}
Livio, M., \& Pringle, J.E. 1994, \apj,  427, 956
\bibitem[Long et al.(1994)]{L94}
Long, K.S., Wade, R.A., Blair, W.P., Davidsen, A.F., \& Hubeny, I. 1994, \apj, 426, 704
\bibitem[Long \& Knigge (2002)]{lk02}
Long, Knox S., \& Knigge, Christian 2002, in Physics of Cataclysmic Variables and Related Objects,
ASP Conference Series vol. 261 eds. B.T. G\"{a}nsicke, K. Beuermann, \& K. Reinsch
(San Francisco:Astronomicasl Society of the Pacific), p.327
\bibitem[Matthews \& Sandage (1963)]{ms63}
Matthews, T.A., \& Sandage, A.R. 1963, \apj, 138, 30
\bibitem[Meyer \& Meyer-Hofmeister (1989)]{MMH89}
Meyer, F., \& Meyer-Hofmeister, E. 1989, A\&A, 221, 36
\bibitem[Meyer \& Meyer-Hofmeister (1994)]{MMH94}
Meyer, F., \& Meyer-Hofmeister, E. 1994, A\&A, 288, 175
\bibitem[Orosz \& Wade (2003)]{ow03}
Orosz, J.A., \& Wade, R.A.  2003, \apj, 593, 1032 
\bibitem[Patterson (1984)]{p1984}
Patterson, J. 1984, \apjs, 54, 443
\bibitem[Pereya et al. (2004)]{p04}
Pereya, N.A., Owocki, S.P., Hillier, D.J., \& Turnshek, D.A. 2004, \apj, 608, 454
\bibitem[Proga, Stone, \& Drew (1998)]{psd98}
Proga, D., Stone, J.M., \& Drew, J.E. 1998, \mnras, 295, 595
\bibitem[Ritter (1988)]{r1988}
Ritter, H. 1988, A\&A, 202, 93
\bibitem[Robinson et al.(1981)]{r1981}
Robinson, E.L., Barker, E.S., Cochran, A.L., Cochran, W.D., \& Nather, R.E.
1981, \apj, 251, 611
\bibitem[Rosino, Romano \& Marziani (1993)]{r1993}
Rosino, L., Romano, G., \& Marziani, P. 1993, \pasp, 105, 51 (RRM)
\bibitem[Rutten, van Paradijs, \& Tinbergen (1992)]{rpt92}
Rutten, R.G.M., van Paradijs, J., \& Tinbergen, J. 1992, \aap, 260, 213
\bibitem[Schenker \& King (2002)]{sk02}
Schenker, K., \& King, A.R. 2002, in The Physics of Cataclysmic Variables and Related Objects,
ASP Conference Series, Vol. 261,
eds. B.T. G\"{a}nsicke, K. Beuermann, \& K. Reinsch 
(San Francisco:Astronomical Society of the Pacific), p.242
\bibitem[Schneider, Young, \& Shectman(1981)]{sch81}
Schneider, D.P., Young, P. \& Shectman, S.A. 1981, \apj, 245, 644
\bibitem[Schreiber, G\"{a}nsicke, \& Cannizzo (2000)]{sgc2000}
Schreiber, M.R., G\"{a}nsicke, B.T., \& Cannizzo, J.K. 2000, \aap, 362, 268
\bibitem[Shakura \& Sunyaev (1973)]{SS73}
Shakura, N.I.,\& Sunyaev, R.A. 1973, A\&A, 24, 337
\bibitem[Sion (1995)]{si95}
Sion, E.M. 1995, \apj, 438, 876
\bibitem [Skillman, Patterson, \& Thorstensen (1995)]{spt95}
Skillman, D.R., Patterson, J., \& Thorstensen, J.R. 1995, \pasp, 107, 545(SPT95)
\bibitem[Smak (1994)]{sm94}
Smak, J. 1994, Acta Astron., 44, 265
\bibitem[Stone et al. (1996)]{s96}
Stone, J.M., Hawley, J.F., Gammie, C.F., \& Balbus, S.A. 1996, \apj, 463, 656
\bibitem[Szkody \& Downes (1982)]{sd82}
Szkody, P., \& Downes, R.A. 1982, \pasp, 94, 328
\bibitem[Townsley \& Bildsten (2002)]{tb02}
Townsley, D.M., \& Bildsten, L. 2002, The Physics of Cataclysmic Variables
and Related Objects, ASP Conference Series, v.261 eds. B.T. G\"{a}nsicke, K. Beuerman, 
\& K. Reinsch (San Francisco:Astronomical Society of the Pacific), p.31
\bibitem[Visniac \& Diamond (1989)]{vd89}
Visniac, E.T., \& Diamond, P. 1989, \apj, 347, 435
\bibitem[Visniac \& Diamond (1992)]{vd92}
Visniac, E.T., \& Diamond, P. 1992, \apj, 398, 561
\bibitem[Warner (1995)]{w95}
Warner, B. 1995, Cataclysmic Variable Stars (Cambridge:Cambridge University Press)
\end{thebibliography}
\end{document}